\documentclass[preprint,aps,prd,nofootinbib,showpacs,superscriptaddress,11pt]{revtex4}
\usepackage{txfonts}
\usepackage[colorlinks=true, linkcolor=red, citecolor=blue]{hyperref}
\usepackage{float}
\usepackage{graphicx}
\usepackage{dcolumn}
\usepackage{bm}
\usepackage{amssymb}
\usepackage{latexsym}
\usepackage{txfonts}
\usepackage{placeins}
\usepackage{natbib}
\usepackage{ifthen}

\newcommand{\be}{\begin{equation}}
\newcommand{\ee}{\end{equation}}
\newcommand{\bq}{\begin{eqnarray}}
\newcommand{\eq}{\end{eqnarray}}

\newbox\tablebox    \newdimen\tablewidth
\def\kmsmpc{\,{\rm km}\,{\rm s}^{-1}\,{\rm Mpc}^{-1}}
\def\leaderfil{\leaders\hbox to 5pt{\hss.\hss}\hfil}

\def\endPlancktable{\tablewidth=\columnwidth
    $$\hss\copy\tablebox\hss$$
    \vskip-\lastskip\vskip -2pt}

\def\tablenote#1 #2\par{\begingroup \parindent=0.8em
    \abovedisplayshortskip=0pt\belowdisplayshortskip=0pt
    \noindent
    $$\hss\vbox{\hsize\tablewidth \hangindent=\parindent \hangafter=1 \noindent
    \hbox to \parindent{$^#1$\hss}\strut#2\strut\par}\hss$$
    \endgroup}
\def\doubleline{\vskip 3pt\hrule \vskip 1.5pt \hrule \vskip 5pt}

\begin{document}

\title{Planck Constraints on Holographic Dark Energy}

\author{Miao Li}
\email{mli@itp.ac.cn} \affiliation{Institute of Theoretical Physics,
Chinese Academy of Sciences, Beijing 100190, China}
\affiliation{Kavli Institute for Theoretical Physics China, Chinese
Academy of Sciences, Beijing 100190, China} \affiliation{Key
Laboratory of Frontiers in Theoretical Physics, Chinese Academy of
Sciences, Beijing 100190, China}

\author{Xiao-Dong Li}
\email{xiaodongli@kias.re.kr} \affiliation{Korea Institute for
Advanced Study, Hoegiro 87, Dongdaemun-Gu, Seoul 130-722, Republic of Korea}

\author{Yin-Zhe Ma}
\email{mayinzhe@phas.ubc.ca} \affiliation{Department of Physics and
Astronomy, University of British Columbia, Vancouver, V6T 1Z1, BC
Canada} \affiliation{Canadian Institute for Theoretical
Astrophysics, 60 St. George Street Toronto, M5S 3H8, Ontario,
Canada}

\author{Xin Zhang}
\email{zhangxin@mail.neu.edu.cn} \affiliation{College of Sciences,
Northeastern University, Shenyang 110004, China}\affiliation{Center
for High Energy Physics, Peking University, Beijing 100080, China}

\author{Zhenhui Zhang}
\email{zhangzhh@itp.ac.cn} \affiliation{Institute of
Theoretical Physics, Chinese Academy of Sciences, Beijing 100190,
China} \affiliation{Kavli Institute for Theoretical Physics China, Chinese
Academy of Sciences, Beijing 100190, China} \affiliation{Key
Laboratory of Frontiers in Theoretical Physics, Chinese Academy of
Sciences, Beijing 100190, China}

\begin{abstract}

We perform a detailed investigation on the cosmological
constraints on the holographic dark energy (HDE) model by using
the {\it Planck} data. We find that HDE can provide a good fit to
the {\it Planck} high-$\ell$ ($\ell\gtrsim 40$) temperature power
spectrum, while the discrepancy at $\ell\simeq 20-40$ found in the
$\Lambda$CDM model remains unsolved in the HDE model. The {\it
Planck} data alone can lead to strong and reliable constraint on
the HDE parameter $c$. At the 68\% confidence level (CL), we
obtain $c=0.508\pm0.207$ with {\it Planck}+WP+lensing, favoring
the present phantom behavior of HDE at the more than 2$\sigma$ CL.
By combining {\it Planck}+WP with the external astrophysical data
sets, i.e. the BAO measurements from 6dFGS+SDSS DR7(R)+BOSS DR9,
the direct Hubble constant measurement result ($H_0=73.8\pm2.4
\kmsmpc$) from the {\it {\it HST}}, the SNLS3 supernovae data set,
and Union2.1 supernovae data set, we get the 68\% CL constraint
results $c=0.484\pm0.070$, $0.474\pm0.049$, $0.594\pm0.051$, and
$0.642\pm0.066$, respectively. The constraints can be improved by
2\%--15\% if we further add the {\it Planck} lensing data into the
analysis. Compared with the {\it WMAP}-9 results, the {\it Planck}
results reduce the error by 30\%--60\%, and prefer a phantom-like
HDE at higher significant level. We also investigate the tension
between different data sets. We find no evident tension when we
combine {\it Planck} data with BAO and {\it {\it HST}}.
Especially, we find that the strong correlation between
$\Omega_{\rm m} h^3$ and dark energy parameters is helpful in
relieving the tension between the {\it Planck} and {\it {\it HST}}
measurements. The residual value of $\chi^2_{Planck\rm +WP+{\it
{\it HST}}}-\chi^2_{Planck\rm +WP}$ is 7.8 in the $\Lambda$CDM
model, and is reduced to 1.0 or 0.3 if we switch the dark energy
to $w$ model or the holographic model. When we introduce
supernovae data sets into the analysis, some tension appears. We
find that the SNLS3 data set is in tension with all other data
sets; for example, for the {\it Planck}+WP, {\it WMAP}-9 and
BAO+{\it {\it HST}}, the corresponding $\Delta \chi^2$ is equal to
6.4, 3.5 and 4.1, respectively. As a comparison, the Union2.1 data
set is consistent with these three data sets, but the combination
Union2.1+BAO+{\it {\it HST}} is in tension with {\it
Planck}+WP+lensing, corresponding to a large $\Delta \chi^2$ that
is equal to 8.6 (1.4\% probability).
Thus, combining internal inconsistent data sets
(SNIa+BAO+{\it HST} with \textit{Planck}+WP+lensing) can lead to ambiguous results,
and it is necessary to perform the HDE data analysis for each independent data sets.
Our tightest self-consistent constraint is $c=0.495\pm0.039$ obtained from
{\it Planck}+WP+BAO+{\it {\it HST}}+lensing.
\end{abstract}


\maketitle

\section{Introduction}

Since the discovery of the cosmic acceleration \cite{Riess}, dark
energy has become one of the most important research areas in
modern cosmology \cite{DEReview}. From the last decade, although a
variety of dark energy models have been proposed to explain the
reason of cosmic acceleration, the physical nature of dark energy
is still a mystery.

The dark energy problem may be in essence an issue of quantum
gravity \cite{Witten:2000zk}. It is commonly believed that the
holographic principle is a fundamental principle of quantum
gravity \cite{Holography}. Based on the effective quantum field
theory, Cohen {\it et al.} \cite{Cohen} suggested that quantum
zero-point energy of a system with size $L$ should not exceed the
mass of a black hole with the same size, i.e., $L^3 \Lambda^4\leq
L M^2_{\rm Pl}$ (here $M_{\rm Pl}\equiv 1/\sqrt{8\pi G}$ is the
reduced Planck mass, and $\Lambda$ is the ultraviolet (UV) cutoff
of the system). In this way, the UV cutoff of a system is related
to its infrared (IR) cutoff. When we consider the whole universe,
the vacuum energy related to this holographic principle can be
viewed as dark energy, and therefore the holographic dark energy
density becomes
\begin{equation}\label{eq:rhohde}
 \rho_{\rm de}=3c^2M^2_{\rm Pl}L^{-2},
\end{equation}
where $c$ is a dimensionless model parameter which modulates the
dark energy density \cite{Li1}. In \cite{Li1}, Li suggested that
the IR length-scale cutoff should be chosen as the size of the
future event horizon of the universe, i.e.,
\begin{equation}\label{eq:Rh}
 L=a\int_t^{+\infty}\frac{dt}{a}.
\end{equation}
This leads to such an equation of state of dark energy
\begin{equation}\label{eq:hdew}
 w_{\rm de}(z)=\frac{1}{3}-\frac{2}{3c}\sqrt{\Omega_{\rm de}(z)},
\end{equation}
which satisfies $w_{\rm de}\approx-0.9$ for $\Omega_{\rm de}=0.7$
and $c=1$. Thus, an accelerated expanding universe can be realized
in this model. In Eq.~(\ref{eq:hdew}), the function $\Omega_{\rm
de}(z)$ is determined by the following coupled differential
equation system
\begin{equation}
\label{eq:deq1}{1\over E(z)}{dE(z) \over dz} =-{\Omega_{\rm
de}(z)\over 1+z}\left({1\over c}\sqrt{\Omega_{\rm
de}(z)}+{1\over2}-{\Omega_{\rm r}(z)+3\over2\Omega_{\rm de}(z)}
\right),
\end{equation}
\begin{equation}
\label{eq:deq2} {d\Omega_{\rm de}(z)\over dz}= -{2\Omega_{\rm
de}(z)(1-\Omega_{\rm de}(z))\over 1+z}\left({1\over
c}\sqrt{\Omega_{\rm de}(z)}+{1\over2}+{\Omega_{r}(z)\over
2(1-\Omega_{\rm de}(z))}\right),
\end{equation}
where $E(z)\equiv H(z)/H_0$ is the dimensionless Hubble expansion
rate, and $\Omega_{\rm r}(z)=\Omega_{\rm r}(1+z)^4/E(z)^2$. Note
that in this paper we only consider a spatially flat universe. The
initial conditions are $E(0)=1$ and $\Omega_{\rm
de}(0)=1-\Omega_{\rm c}-\Omega_{\rm b}-\Omega_{\rm r}$.


The holographic dark energy (HDE) model described above is a
viable and physically plausible dark energy candidate, as an
alternative to the standard cosmological constant model
($\Lambda$). The model has been widely studied both theoretically
\cite{HDEworks} and observationally \cite{HDEObserv}. The data
used in these works mainly include the type Ia supernovae (SNIa),
baryon acoustic oscillations (BAO), the direct measurement of
Hubble constant, and the Cosmic Microwave Background (CMB) data
from the {\it Wilkinson Microwave Anisotropy Probe} ({\it WMAP}).
These works show that the HDE model can provide a good fit to the
data, and $c<1$ is favored by the data. For example, a recent
analysis reports the 68\% confidence level (CL) constraint
$c=0.680^{+0.064}_{-0.066}$ from {\it WMAP}-7+SNIa+BAO+{\it {\it
HST}}~\cite{HDEGlobal}.

In this March, the European Space Agency (ESA) and the {\it
Planck} Collaboration publicly released the CMB data based on the
first 15.5 months of {\it Planck} operations, along with a lot of
scientific results \cite{Planck01}. They show that the standard
six-parameter $\Lambda$CDM model provides an extremely good fit to
the {\it Planck} spectra at high multipoles, while there are some
discrepancy at $\ell\simeq20-40$. Some cosmological parameters,
e.g., $n_{\rm s}$, $\Omega_{\rm k}$, and $N_{\rm eff}$, are
measured with unprecedented precision. Interestingly, the {\it
Planck} values for some $\Lambda$CDM parameters are significantly
different from those previously measured. For the matter density
parameter, the {\it Planck} data give $\Omega_{\rm
m}=0.315\pm0.017$ (68\% CL) \cite{Planck16}. This value is higher
than the {\it WMAP}-7 result $\Omega_{\rm m}=0.273\pm0.030$
\cite{WMAP7} and the {\it WMAP}-9 result $\Omega_{\rm
m}=0.279\pm0.025$ \cite{WMAP9}, and is in tension with the SNLS3
result $\Omega_{\rm m}=0.211 \pm 0.069$ \cite{SNLS3}. For the
Hubble constant, \textit{Planck} gives a low value $H_0=67.3\pm1.2
\kmsmpc$, which is in tension with the results of the direct
measurements of $H_0$, i.e., $H_0=73.8\pm2.4 \kmsmpc$ reported by
Riess {\it et al.} \cite{HSTWFC3}, and $H_0=74.3\pm1.5\ ({\rm
statistical}) \pm2.1\ ({\rm systematic}) \kmsmpc$ reported by
Freedman {\it et al.} \cite{CarnegieHP}. The discrepancy is at
about the 2.5$\sigma$ level. They also show that, the {\it Planck}
constraints of $\Omega_{\rm m}$ and $H_0$, although are in tension
with SNLS3 and {\it HST} observations, are in agreement with the
geometrical constraints from BAO surveys \cite{Planck16}.

The {\it Planck} data also improve the constraints on dark energy
\cite{Planck16}. Actually, the results can be significantly
different if the {\it Planck} data are combined with different
astrophysical data sets. For a constant $w$ model (here after,
$w$CDM model), the {\it Planck} results give
$w=-1.13^{+0.13}_{-0.10}$ and $w=-1.09\pm0.17$ (95\% CL) by using
CMB combined with BAO and Union2.1 \cite{Union2p1} data,
respectively, which are consistent with the cosmological constant.
However, when combined with SNLS3 data and $H_0$ measurement, the
results are $w=-1.13^{+0.13}_{-0.14}$ and
$w=-1.24^{+0.18}_{-0.19}$ ($2\sigma$ CL), respectively, favoring
$w<-1$ at the 1--2$\sigma$ level. For a dynamical equation of state
$w=w_0+w_a(1-a)$, the results from the {\it Planck}+WP+BAO and
{\it Planck}+WP+Union2.1 data combinations are in agreement with a
cosmological constant, while the {\it Planck}+WP+$H_0$ and {\it
Planck}+WP+SNLS3 (here, WP represents the {\it WMAP}-9
polarization data) results are in tension with $w=-1$ at the more
than 2$\sigma$ level.

Based on the arrival of a bunch of new data sets, it is very
important to re-analyze the HDE model in light of {\it Planck} and
\textit{WMAP} 9-year data. This will enable us to answer a lot of
interesting questions: What are the constraint results of the
cosmological parameters in the HDE model from the {\it Planck}
data? What is the difference between the fitting results of {\it
Planck} and {\it WMAP}? What are the results if we combine the
{\it Planck} data with the BAO, SNIa, and {\it HST} data? Whether
are they consistent or in tension with each other? Since the
Hubble constant $H_0$ is correlated with the HDE parameter $c$,
can HDE help us to relax the tension between the {\it Planck} data
and the direct measurements of $H_0$? Since a phantom dark energy
can reduce the TT power spectrum amplitude at large scales, can
HDE help us to relieve the mismatches between theoretical and
observational power spectra at $\ell\simeq20-40$?
The $Planck$ temperature power spectrum showed anomalous fitting results 
of the lensing parameter $A_{\rm L}$ in the $\Lambda$CDM model (i.e., $A_{\rm L}>1$),
can HDE help us to remove or relieve this ``anomaly''?
The main task of this paper is to find firm, reliable answers to these stimulating questions.

This paper is organized as follows. In Sec.~\ref{sec:data}, we
give a brief introduction to the data used in this work and our
method of data analysis. In Sec.~\ref{sec:cmb}, we present and
compare the fitting results of HDE by using the CMB-only data of
{\it Planck} and {\it WMAP}-9. In Sec.~\ref{sec:otherdata}, we
combine the CMB data with the external astrophysical data sets
including BAO, SNLS3, Union2.1 and {\it HST}, and discuss the
fitting results and the tensions. Some concluding remarks are
given in Sec.~\ref{sec:concl}. In this work, we assume today's
scale factor $a_0 = 1$, so the redshift $z$ satisfies $z = 1/a -
1$. We use negative redshifts to represent the future; in this
way, $z=-1$ corresponds to the infinite future when
$a\rightarrow\infty$. The subscript ``0'' indicates the present
value of the corresponding quantity unless otherwise specified.

\section{Data analysis methodology}\label{sec:data}


To analyze the HDE, we modify the \texttt{CAMB} package
\cite{CAMB} to incorporate the background equations of the HDE model. 
Furthermore, to investigate the dark energy perturbations, we
apply the ``parameterized post-Friedmann'' (PPF) approach
\cite{PPF}. This method of dealing with dark energy perturbations
has been widely used by {\it WMAP} ~\cite{WMAP7,WMAP9} and {\it
Planck} teams~\cite{Planck16}. In our previous work of HDE data
analysis \cite{HDEGlobal}, we have already employed this method
into our pipeline.

The same as \cite{Planck16}, we sample cosmological parameter
space with Markov Chain Monte Carlo (MCMC) method with the
publicly available code \texttt{COSMOMC} \cite{COSMOMC}. For each
analysis, we execute about 8--16 chains until they are converged,
satisfying the standard Gelman and Rubin criterion $R-1<0.01$
\cite{GRCri}. To make sure that the tails of the distribution are
well enough explored, we also check the convergence of confidence
limits with the setting \texttt{MPI\_Limit\_Converge} = 0.025 in
\texttt{COSMOMC}.

The base $\Lambda$CDM model has the standard ``six-parameter'' as
\begin{equation}
 {\rm P}=\{\Omega_{\rm b}h^2,~\Omega_{\rm c}h^2,~100\theta_{\rm MC},~\tau,~n_{\rm s},~\ln(10^{10}A_{\rm s})\},
\end{equation}
where $\Omega_{\rm b}h^2$ and $\Omega_{\rm c}h^2$ are the current
density of baryon and cold dark matter, respectively,
$100\theta_{\rm MC}$ is $100$ times the approximation to $r_{\rm
s}/D_{\rm A}$ in \texttt{COSMOMC} ($r_{\rm s}=r_{\rm s}(z_{\rm
drag})$ is the comoving size of sound horizon at baryon-drag
epoch, and $D_{\rm A}$ is the angular diameter distance), $\tau$
is the Thomson scattering optical depth due to reionization,
$n_{\rm s}$ is the scalar spectrum index at the pivot scale
$k_0=0.05\ {\rm Mpc}^{-1}$, $\ln(10^{10}A_{\rm s})$ is the log
power of the primordial curvature perturbations at $k_0$.

In the following, we will also discuss the holographic dark energy
model and the $w$CDM model, each of which has an extra parameter
to describe the dynamic evolution of dark energy. For HDE model,
the extra parameter is $c$, as described in Eq.~(\ref{eq:rhohde}),
and for $w$CDM model, the extra parameter is $w$. Therefore, when
we compare $\Lambda$CDM model with $w$CDM and HDE models, we should
bear in mind that we are comparing a model with $6$ parameters
with models with $7$ parameters.

To make our results comparable with the results of the {\it
Planck} Collaboration, baselines and priors for the parameters in
our analysis are adopted same as \cite{Planck16}. In our MCMC
chains, these parameters are varied with uniform priors, within
the ranges listed in Table 1 of \cite{Planck16}. The range of $c$
is [0.001,~3.5], which is wide enough for covering the physically
interesting region. Additionally, a ``hard'' prior $[20,~100]
\kmsmpc$ is imposed to the Hubble constant
\footnote{In the MCMC, samples with $H_0$ out of this range are rejected.}.
The same as
\cite{Planck16}, we assume a minimal-mass normal hierarchy for the
neutrino masses by setting a single massive eigenstate
$m_{\nu}=0.06$ eV.

Cosmological data used in this work fall into two parts: the CMB
data from {\it Planck} and {\it WMAP}, and the other data sets
including BAO, SNIa and $H_0$. We introduce them in the following
two subsections.

\subsection{CMB data}

The CMB data based on the first 15.5 months of {\it Planck}
operations are publicly released by the ESA and {\it Planck}
Collaboration in March 2013 \cite{Planck01}. At the same time, the
{\it Planck} likelihood softwares are also made publicly
downloadable.
\footnote{http://pla.esac.esa.int/pla/aio/planckProducts.html} The
likelihood software provided by the {\it Planck} Collaboration
includes the following four parts:
\begin{itemize}
 \item The high-$\ell$ temperature likelihood \texttt{CamSpec}.
 At $\ell=50-2500$, a correlated Gaussian approximation is employed to obtain the likelihood,
 based on a fine-grained set of angular cross-spectra derived from multiple detector combinations between
 the 100, 143, and 217 GHz frequency channels.
 \item The low-$\ell$ temperature likelihood.
 At $\ell<50$, the likelihood exploits all {\it Planck} frequency channels from 30-353 GHz,
 separating the CMB signal from the diffuse Galactic foregrounds through a physically motivated Bayesian component separation technique.
 \item The low-$\ell$ polarization likelihood.
 The present {\it Planck} data release includes only temperature data,
 and the {\it Planck} Collaboration supplements the {\it Planck} likelihood with the 9-year {\it WMAP} ({\it WMAP}-9) polarization likelihood derived
 from the {\it WMAP} polarization maps at 33, 41, and 61 GHz (K, Q, and V bands).
 \item The {\it Planck} lensing likelihood.
 Lensing is detected independently in {\it Planck} 100, 143, and 217 GHz channels with an overall significance of greater than 25$\sigma$ \cite{Planck17}.
 The gravitational lensing data are good at constraining dark energy through the lensing effect coming from the distortion
 of the large scale structure that emerged after $z=10$ (at this stage, the universe is dark energy dominated).
\end{itemize}
In the following context, we will use ``{\it Planck}'' to
represent the {\it Planck} temperature likelihood (including both
the low-$\ell$ and high-$\ell$ parts), ``WP'' to represent the
{\it WMAP} polarization likelihood as a supplement of {\it
Planck}, and ``lensing'' to represent the likelihood of {\it
Planck} lensing data.

To study the difference between the fitting results by using {\it
Planck} and {\it WMAP} data, in this work we also perform the
analysis of HDE by using {\it WMAP}-9 data. The data and likelihood
software are downloadable at the Legacy Archive for Microwave
Background Data Analysis (LAMBDA).
\footnote{http://lambda.gsfc.nasa.gov} We will not use the
high-resolution CMB data of the Atacama Cosmology Telescpoe and the
South Pole Telescope \cite{ACTSPT}. They are not publicly available
in the current version of \texttt{COSMOMC} package, and only
marginally affect the fitting results compared with {\it Planck} or
{\it WMAP}-9.

\subsection{External astrophysical data sets}

The CMB data alone are not powerful in constraining dark energy
parameters, since dark energy affects the late time cosmic
evolution. When combined with the external astrophysical data sets
(hereafter, ``Ext'' or ``Exts''), CMB data are helpful in breaking
the degeneracies between parameters and improving the constraints
on dark energy parameters \cite{H0WHu}. In our analysis, we will
consider the following four Exts:
\begin{itemize}
 \item
 The BAO data can provide effective constraints on dark energy from the angular diameter distance--redshift relation.
 In our analysis, similar to \cite{Planck16},
 we use the following data sets,
 the 6dF Galaxy Survey $D_{\rm V}(0.106)=(457\pm27)$Mpc \cite{6dFGS} ($D_{\rm V}$ is a distance indicator similar to angular diameter distance $D_{\rm A}$,
 see Eq.~(46) in \cite{Planck16}),
 the reanalyzed SDSS DR7 BAO measurement $D_{\rm V}(0.33)/r_{\rm s}=8.88\pm0.17$ \cite{SDSSDR7R},
 and the BOSS DR9 measurement $D_{\rm V}(0.57)/r_{\rm s}=13.67\pm0.22$ \cite{BOSSDR9}.
 SDSS DR7 and BOSS DR9 are the two most accurate BAO measurements,
 and the correlation between the surveys is a marginal effect to the parameter estimation.
 \item
 The direct measurement of the Hubble constant, $H_0=73.8\pm2.4 \kmsmpc$ ($1\sigma$ CL) \cite{HSTWFC3},
 from the supernova magnitude--redshift relation calibrated by the {\it {\it HST}} observations of Cepheid variables in the host galaxies of eight SNe Ia.
 Here the uncertainty is 1$\sigma$ and includes known sources of systematic errors.
 \item
 The Union2.1 compilation \cite{Union2p1}, consisting of 580 SNe,
 calibrated by the SALT2 light-curve fitting model \cite{SALT2}.
 \item
 The SNLS3 ``combined'' sample \cite{SNLS3}, consisting of 472 SNe,
 calibrated by both SiFTO \cite{SiFTO} and SALT2 \cite{SALT2}.
 For simplicity, we do not consider the SNLS3 compilation calibrated separately by SiFTO or SALT2.
\end{itemize}
In the following context, we will use ``BAO'', ``{\it {\it
HST}}'', ``Union2.1'' and ``SNLS3'' to represent these four Exts.
We will also use ``SNIa'' to represent a supernovae data set,
either Union2.1 or SNLS3.

\section{CMB-only results}\label{sec:cmb}

\begin{table}[!htp]
\caption{\label{Table:CMBonly} CMB-only fitting results of the HDE
model.}
\begingroup
\openup 5pt
\newdimen\tblskip \tblskip=5pt
\nointerlineskip
\vskip -3mm
\scriptsize
\setbox\tablebox=\vbox{
    \newdimen\digitwidth
    \setbox0=\hbox{\rm 0}
    \digitwidth=\wd0
    \catcode`"=\active
    \def"{\kern\digitwidth}
    \newdimen\signwidth
    \setbox0=\hbox{+}
    \signwidth=\wd0
    \catcode`!=\active
    \def!{\kern\signwidth}
\halign{ \hbox to
2.3in{$#$\leaderfil}\tabskip=1.5em&\hfil$#$\hfil&\hfil$#$\hfil&\hfil$#$\hfil&\hfil$#$\hfil&\hfil$#$\hfil&\hfil$#$\hfil\tabskip=0pt&\hfil$#$\hfil\tabskip=0pt\cr
\noalign{\doubleline} \multispan1\hfil \hfil&\multispan2\hfil
$\Omega_{\rm m}$ \hfil&\multispan2\hfil $c$ \hfil&\multispan2\hfil
$H_0$ \hfil\cr \noalign{\vskip -3pt}
\omit&\multispan2\hrulefill&\multispan2\hrulefill&\multispan2\hrulefill&\omit\cr
\omit\hfil Data\hfil&\omit\hfil Best fit\hfil&\omit\hfil 68\%
limits\hfil&\omit\hfil Best fit\hfil&\omit\hfil 68\%
limits\hfil&\omit\hfil Best fit\hfil&\omit\hfil 68\%
limits\hfil&\omit\hfil $\ \ \ \ \ \ -\ln \mathcal{L}_{max}\ \ \ \
\ $\hfil\cr \noalign{\vskip 5pt\hrule\vskip 3pt} Planck{\rm } &
0.142 & 0.261\pm0.097 & 0.301 & 0.587\pm0.449 & 100.00 &
77.34\pm12.82 & 3894.4\cr Planck{\rm +lensing} & 0.150 &
0.248\pm0.084 & 0.317 & 0.531\pm0.296 & 96.80 & 78.51\pm12.09 &
3899.8\cr Planck{\rm +WP} & 0.157 & 0.268\pm0.100 & 0.317 &
0.612\pm0.433 & 95.57 & 75.60\pm12.83 & 4902.6\cr Planck{\rm
+WP+lensing} & 0.180 & 0.248\pm0.079 & 0.354 & 0.508\pm0.207 &
88.65 & 78.36\pm11.36 & 4907.3\cr \noalign{\vskip 5pt\hrule\vskip
3pt} {\rm {\it WMAP}\textendash9} & 0.350 & 0.401\pm0.082 & 0.965
& 1.88_{-1.20}^{+0.79} & 62.37 & 59.57\pm8.15 & 3779.0\cr
\noalign{\vskip 5pt\hrule\vskip 3pt}
} 
} 
\endPlancktable
\endgroup
\end{table}

In this section we present the CMB-only fitting results of the HDE
model. The CMB+Ext fitting results are discussed in the next
section.

In Table~\ref{Table:CMBonly}, we list the fitting results of the
HDE model from the CMB data alone. Best-fit values as well as the
68\% CL limits for $\Omega_{\rm m}$, $c$ and $H_0$ are listed in
columns 2--7. The minus log-maximal likelihood is listed in the
last column. The first 4 rows list the results of {\it Planck},
{\it Planck}+lensing, {\it Planck}+WP, and {\it
Planck}+WP+lensing. For comparison, the {\it WMAP}-9 results are
listed in the last row.

In the following two subsections, we firstly introduce the
temperature power spectra with the best-fit parameters, and then
discuss the constraints on cosmological parameters.

\subsection{Temperature power spectra}

\begin{figure}[H]
 \centering{
\includegraphics[width=16cm]{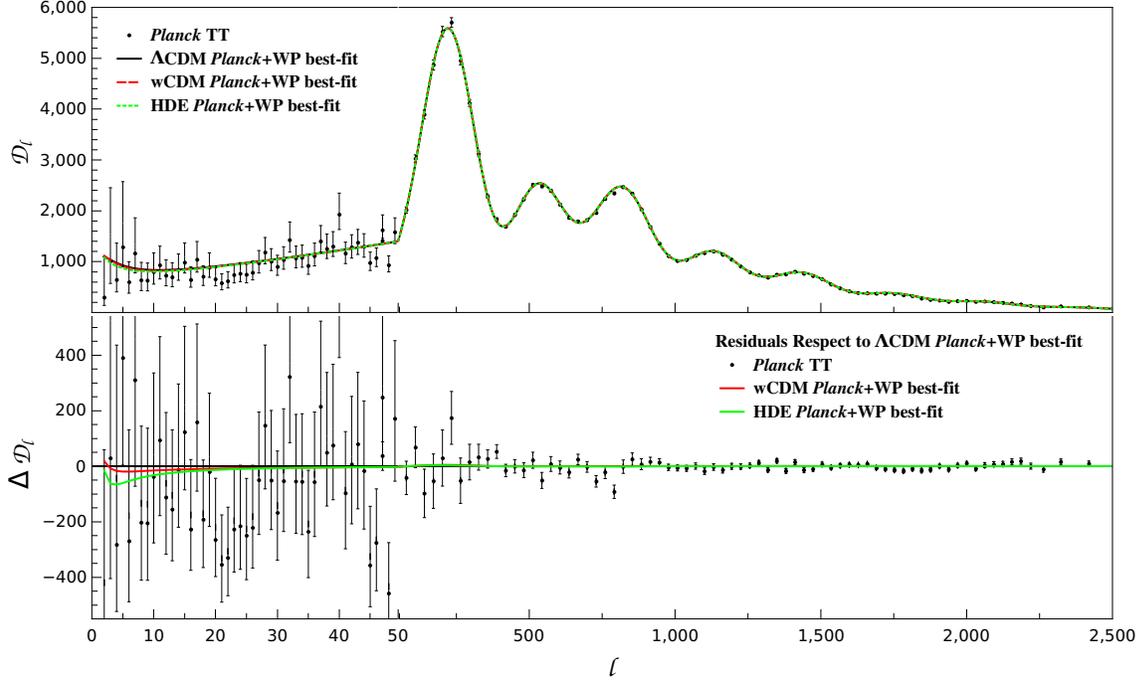}
} \caption{\label{Fig:Dl} Upper panel: CMB TT power spectrum
plotted with the best-fit parameters of $\Lambda$CDM model (black
solid), $w$CDM model (red dashed), and HDE model (green dotted),
from the {\it Planck}+WP data. The ordinate axis shows
$\mathcal{D}_{\ell}\equiv \ell(\ell+1)C_{\ell}/2\pi$ in units of
$\mu \textrm{K}^2$. The {\it Planck} binned temperature spectrum
is shown in black dots with error bars. Lower panel: Residuals
with respect to the temperature power spectrum of the best-fit
six-parameter $\Lambda$CDM model.}
\end{figure}

In the upper panel of Fig.~\ref{Fig:Dl} we show the temperature
power spectrum of the best-fit HDE model (green dotted) by using
the {\it Planck}+WP data. As comparisons, best-fit spectra of the
$\Lambda$CDM model and the $w$CDM model from {\it Planck}+WP are
also plotted in black solid and red dashed lines. To see the
difference between the three spectra, the residuals compared with
the best-fit six-parameter $\Lambda$CDM model are shown in the
lower panel. We find that all the three models can provide a good
fit to the {\it Planck} high-$\ell$ power spectrum, while at
$\ell\simeq20-40$ there are some mismatches, as reported by {\it
Planck} \cite{Planck16}. The HDE model is not helpful in relieving
this discrepancy. The main difference among the power spectra of
the three models lie in the $\ell\lesssim20$ region, where we find
that amplitudes of HDE and $w$CDM spectra are lower than the
$\Lambda$CDM spectrum. This phenomenon is consistent with the
result of \cite{HDEGlobal}, where it is shown that a phantom-like
dark energy component leads to smaller $C_{\ell}^{TT}$ at
low-$\ell$ region.

\begin{figure}[H]
\centering{
\includegraphics[width=16cm]{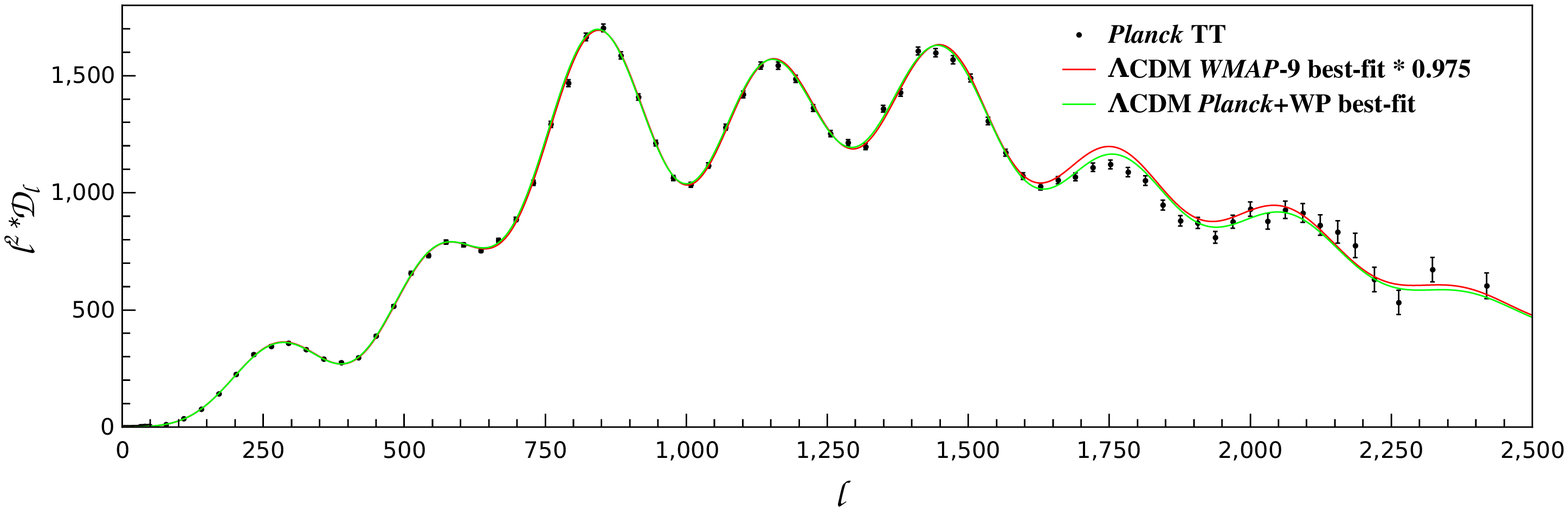}
\includegraphics[width=16cm]{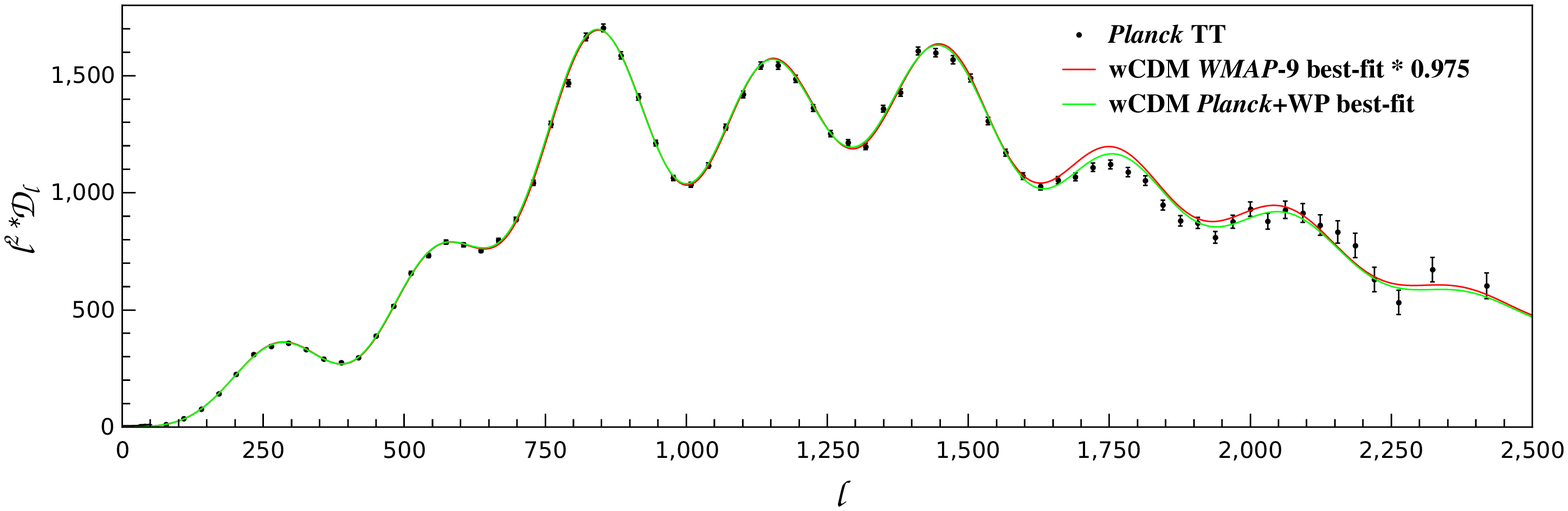}
\includegraphics[width=16cm]{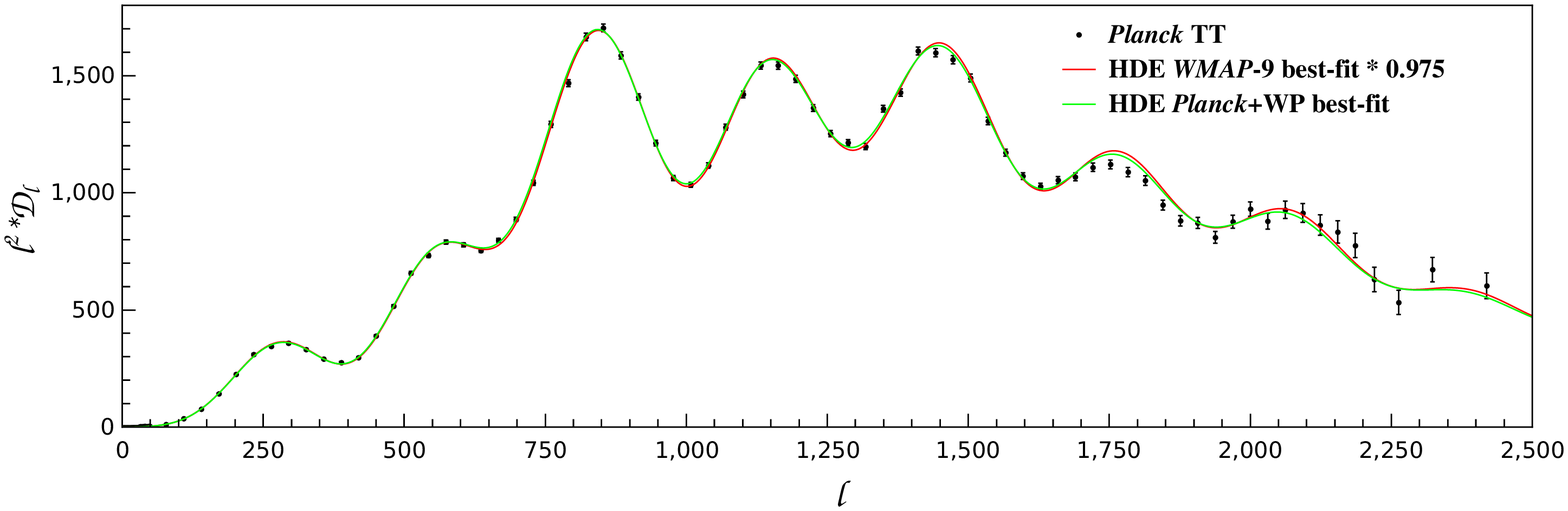}
} \caption{\label{Fig:l2Dl} The {\it WMAP}-9 and {\it Planck}+WP
best-fit power spectra for the $\Lambda$CDM (upper panel), $w$CDM
(middle panel) and HDE (lower panel) models. To see the difference
between the theoretical power spectra and the observational data
at the high-$\ell$ region, we choose to plot the $\ell^2
\mathcal{D}_\ell$ (in units of $\textrm{mK}^{2}$) rather than
$\mathcal{D}_\ell$. The {\it Planck}+WP best-fit power spectra are
plotted in green lines, and the {\it WMAP}-9 best-fit power
spectra multiplied by 0.975 are plotted in red lines. The black
points with error bars mark the {\it Planck} temperature power
spectrum data.}
\end{figure}

It is also of interest to compare the {\it WMAP} and {\it Planck}
spectra in the HDE model. The Appendix A of \cite{Planck16} shows
some inconsistency between the {\it Planck} and {\it WMAP}
spectra. It is found that the {\it WMAP} power spectrum re-scaled
by a multiplicative fator of 0.975 agree to remarkable precision
with the {\it Planck} spectrum~\cite{Planck16}. Thus, in Fig.
\ref{Fig:l2Dl} we plot the {\it WMAP}-9 and {\it Planck}+WP
spectra for the $\Lambda$CDM (upper panel), $w$CDM (middle panel)
and HDE (lower panel) models. As expected, in all these three
models, we find that the {\it WMAP}-9 power spectrum (with a
multiplicative factor 0.975) matches well with the {\it Planck}
power spectrum. The best-fit power spectra of the three models are
similar to each other. More interestingly, in all models we find
that at $\ell\sim 1600-2000$ the theoretical power spectra of {\it
Planck} and {\it WMAP}-9 have higher amplitudes than the {\it
Planck} data. This scale corresponds to $\sim$10 times the scale
of galaxy clusters, and this discrepancy may be due to some
unclear physics on this scale.

\subsection{Constraints on cosmological parameters}

In this subsection we discuss the constraints on cosmological
parameters in the HDE model.

\begin{figure}[H]
\centering{
\includegraphics[width=8cm,height=6cm]{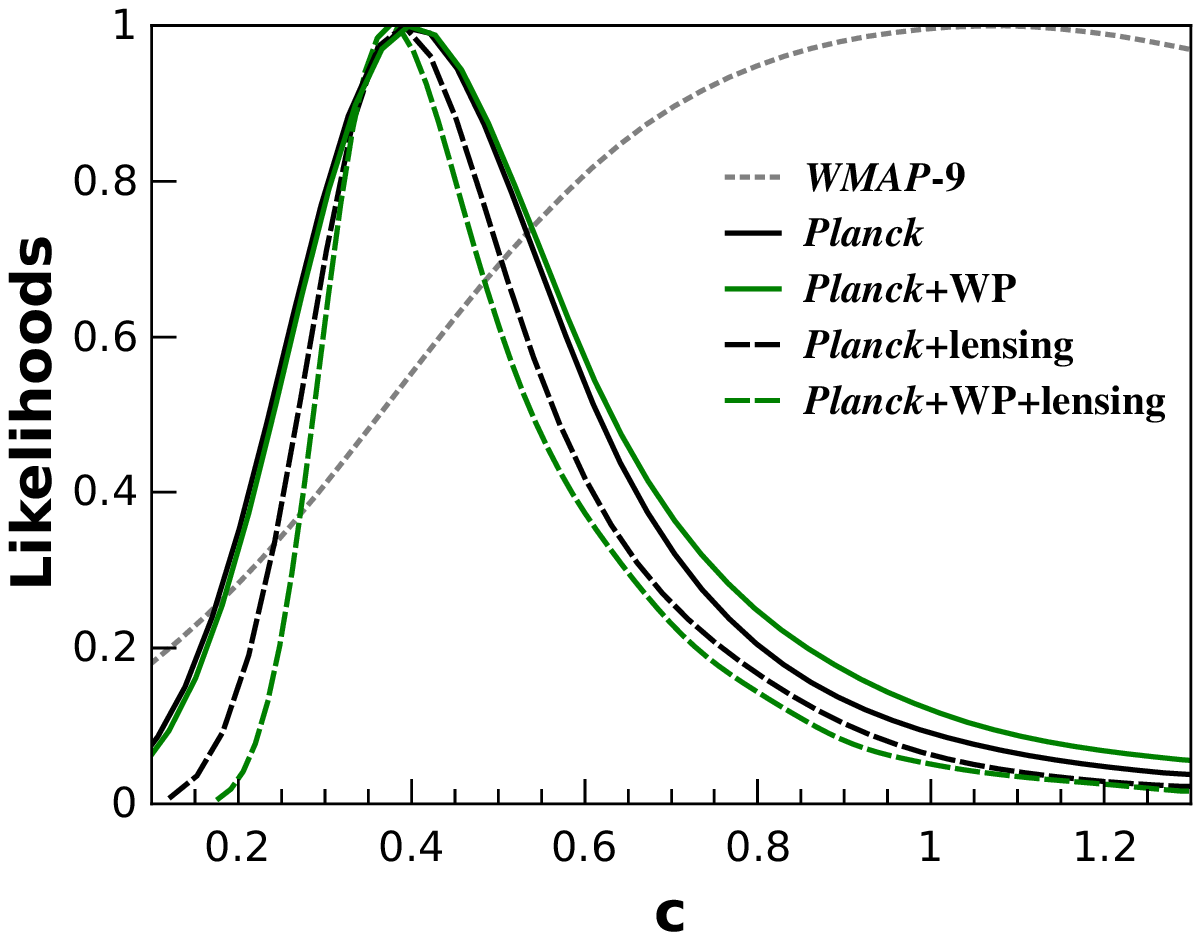}
\includegraphics[width=8cm,height=6cm]{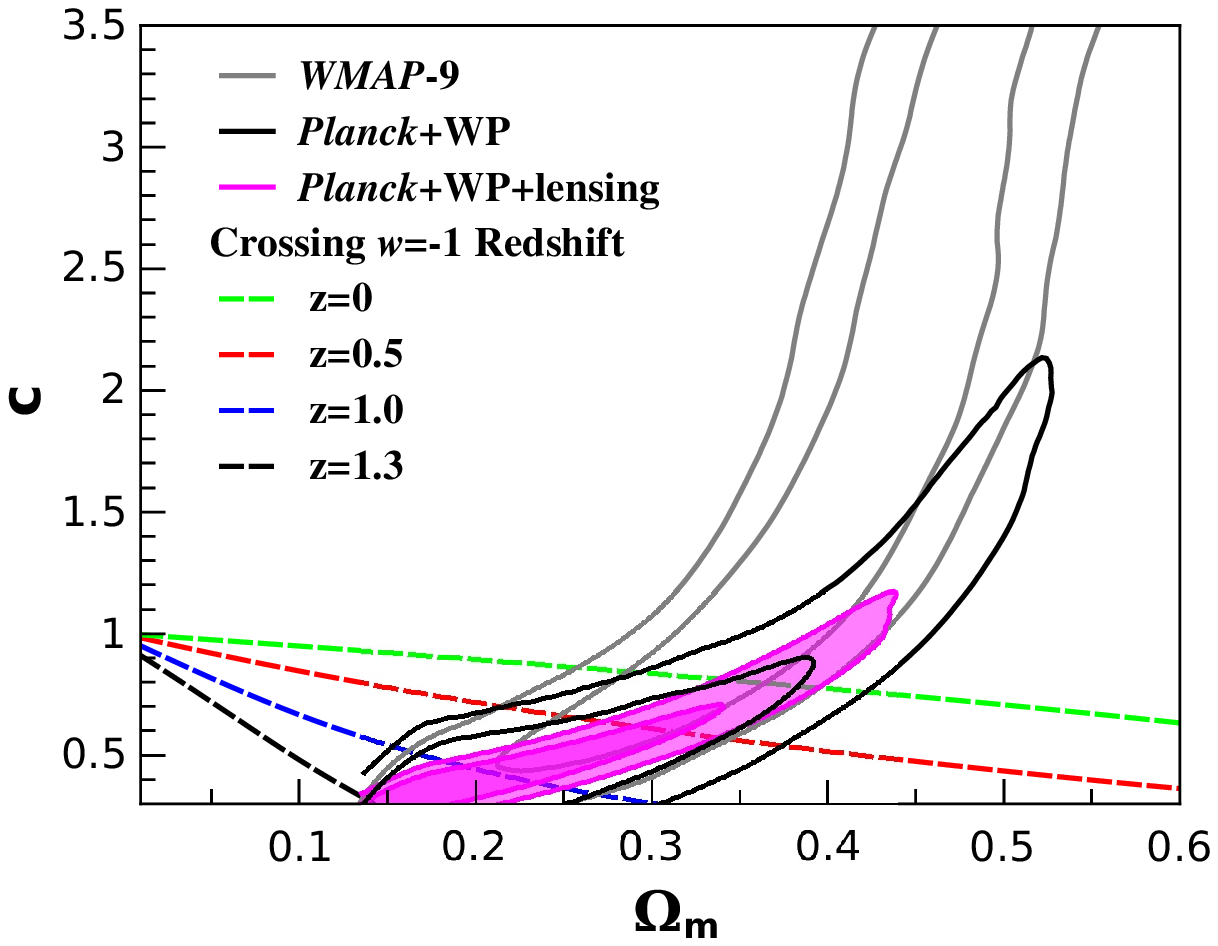}}
\caption{\label{Fig:cmbonly} CMB-only fitting results of the HDE
model. Left panel: Marginalized likelihood distributions of $c$.
Right panel: Marginalized 68\% and 95\% CL contours in the
$\Omega_{\rm m}$--$c$ plane. Dashed lines mark the $w=-1$ crossing
at $z=0,\ 0.5,\ 1.0$ and 1.3.}
\end{figure}

The likelihood distributions of $c$ are shown in the left panel of
Fig. \ref{Fig:cmbonly}. We find that {\it Planck} data lead to small
values of best-fit $c$, i.e., 0.30--0.35. The 68\% CL errors of $c$
are about 0.45 for {\it Planck} and {\it Planck}+WP, and are reduced
to 0.3 and 0.21 when the lensing data are added. Compared with the
{\it WMAP}-9 alone constraint, $c=1.786\pm0.880$, the {\it Planck}
results reduce the error bar by about 45\%--75\%. There are clear
discrepancies between the mean (see the 68\% limits listed in Table \ref{Table:CMBonly}) and the
best-fit values of $c$, implying that the likelihood distribution of
$c$ is highly deviated from symmetric form.

The right panel of Fig.~\ref{Fig:cmbonly} shows the $\Omega_{\rm
m}$--$c$ contours of the CMB-only constraints. Results of {\it
WMAP}-9, {\it Planck}+WP and {\it Planck}+WP+lensing are plotted.
To see the behavior of HDE under the constraints, we also plot the
``crossing $w=-1$ redshift'' in dashed lines: e.g., parameter
space above/bellow the dashed blue line corresponds to a
quitessence/phantom behavior of holographic dark energy at
$z=1.0$. We see that the {\it WMAP}-9 data alone does not lead to
any interesting constraint on $c$, while the {\it Planck}+WP
results show the preference for $c < 1$ at the 1$\sigma$ CL.
Adding the lensing data tightens the constraint, and the present
phantom behavior of holographic dark energy is prefered at the
more than 1$\sigma$ CL. Besides, we find that in the HDE model
$\Omega_{\rm m}$ is constrained to be 0.26--0.28 (68\% CL) by the
{\it Planck} data, which is smaller than the result in the
$\Lambda$CDM model. The {\it WMAP} data alone cannot lead to
effective constraint on $\Omega_{\rm m}$ in the HDE model.

The CMB-only constraints on $H_0$ in the HDE model are listed in the
5$th$ and 6$th$ columns of Table \ref{Table:CMBonly}. Compared with
the $\Lambda$CDM result, $H_0=67.4\pm1.4$ (68\% CL; {\it Planck})
\cite{Planck16}, the error bars are significantly larger.
\footnote{Since we impose a prior [20,~100] on $H_0$ in the
analysis, the error bars are, actually, under-estimated.} Similar
phenomenon appears in the {\it WMAP}-9 results, where we find
$H_0=59.57\pm8.15$ in the HDE model.

\begin{figure}[H]
 \centering{
\includegraphics[width=8cm,height=6cm]{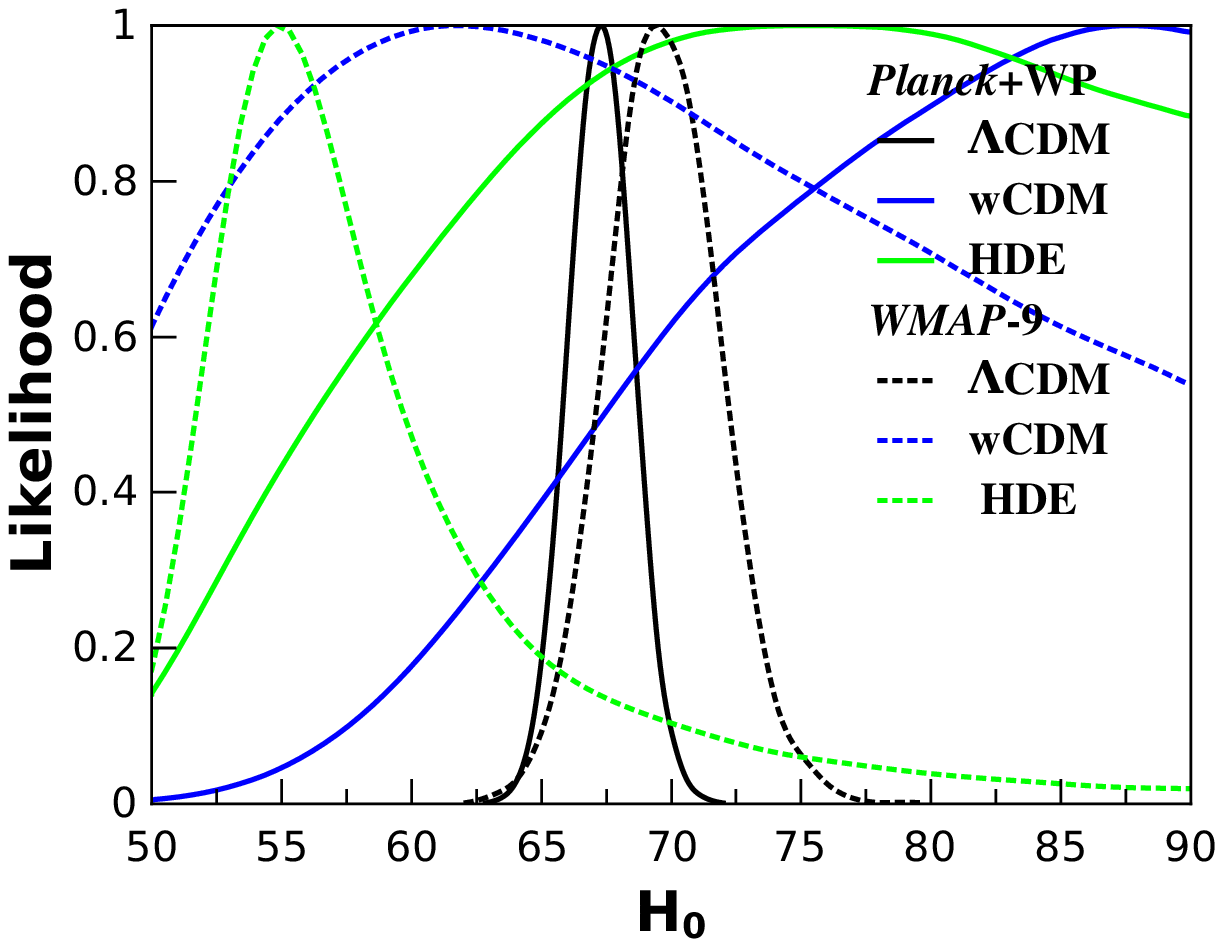}
\includegraphics[width=8cm,height=6cm]{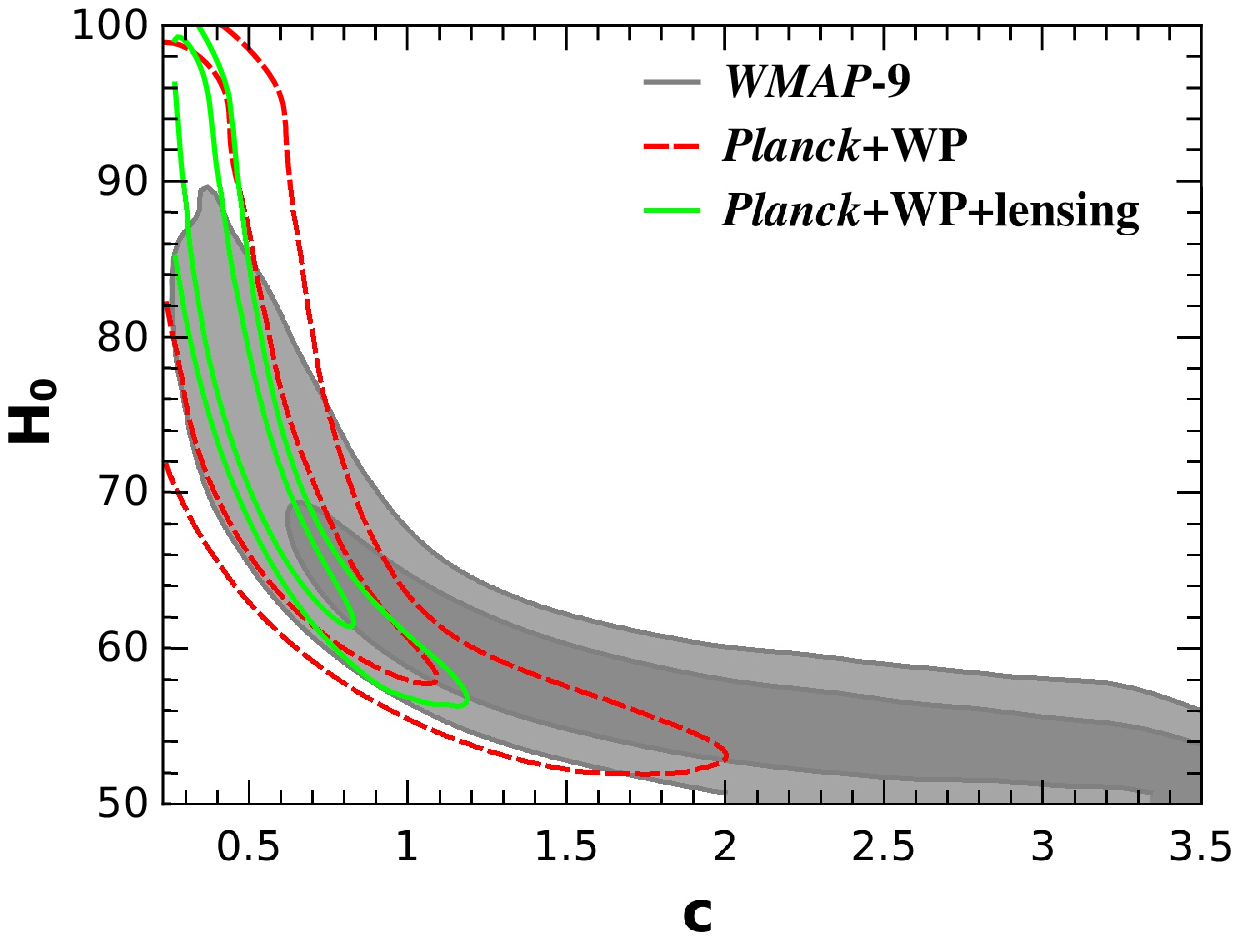}
} \caption{\label{Fig:H0cmb} CMB-only fitting results of the HDE
model. Left panel: Marginalized likelihood distributions of $H_0$.
Right panel: Marginalized 68\% and 95\% CL contours in the
$c$--$H_0$ plane. }
\end{figure}

To make a comparison, in the left panel of Fig. \ref{Fig:H0cmb} we
plot the likelihood distributions of $H_0$ in the $\Lambda$CDM,
$w$CDM and HDE models, constrained by {\it Planck}+WP and {\it
WMAP}-9 data. We find that, in the $\Lambda$CDM model, $H_0$ is
tightly constrained, while in the HDE and $w$CDM models it cannot be
effectively constrained. The right panel shows the $c$--$H_0$
contours constrained by {\it WMAP}-9, {\it Planck}+WP, and {\it
Planck}+WP+lensing. We see that $c$ and $H_0$ are strongly
anti-correlated with each other. This explains why in the HDE model
$H_0$ cannot be well constrained by CMB-only data.

\begin{figure}[H]
\centering{
\includegraphics[width=8cm,height=6cm]{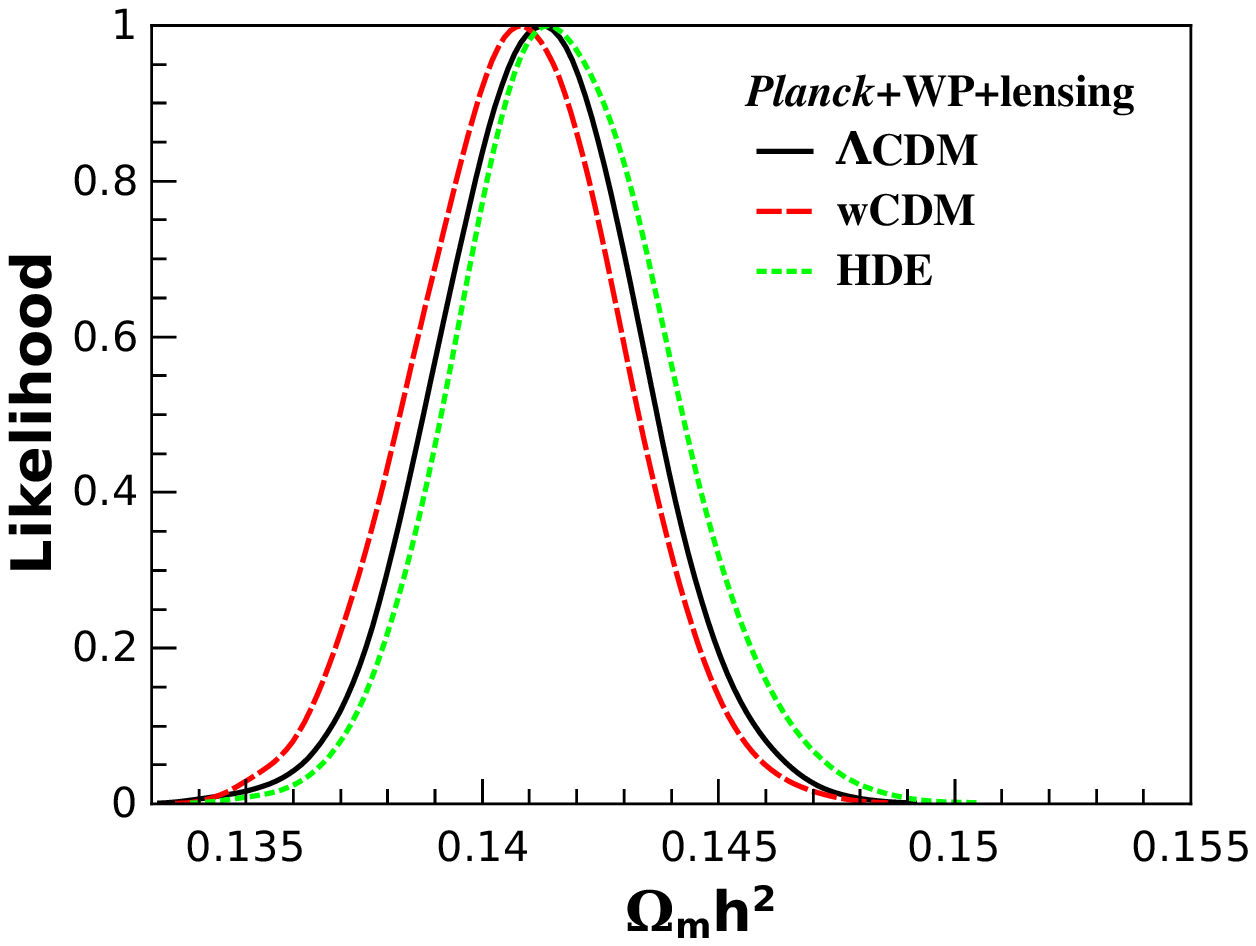}
\includegraphics[width=8cm,height=6cm]{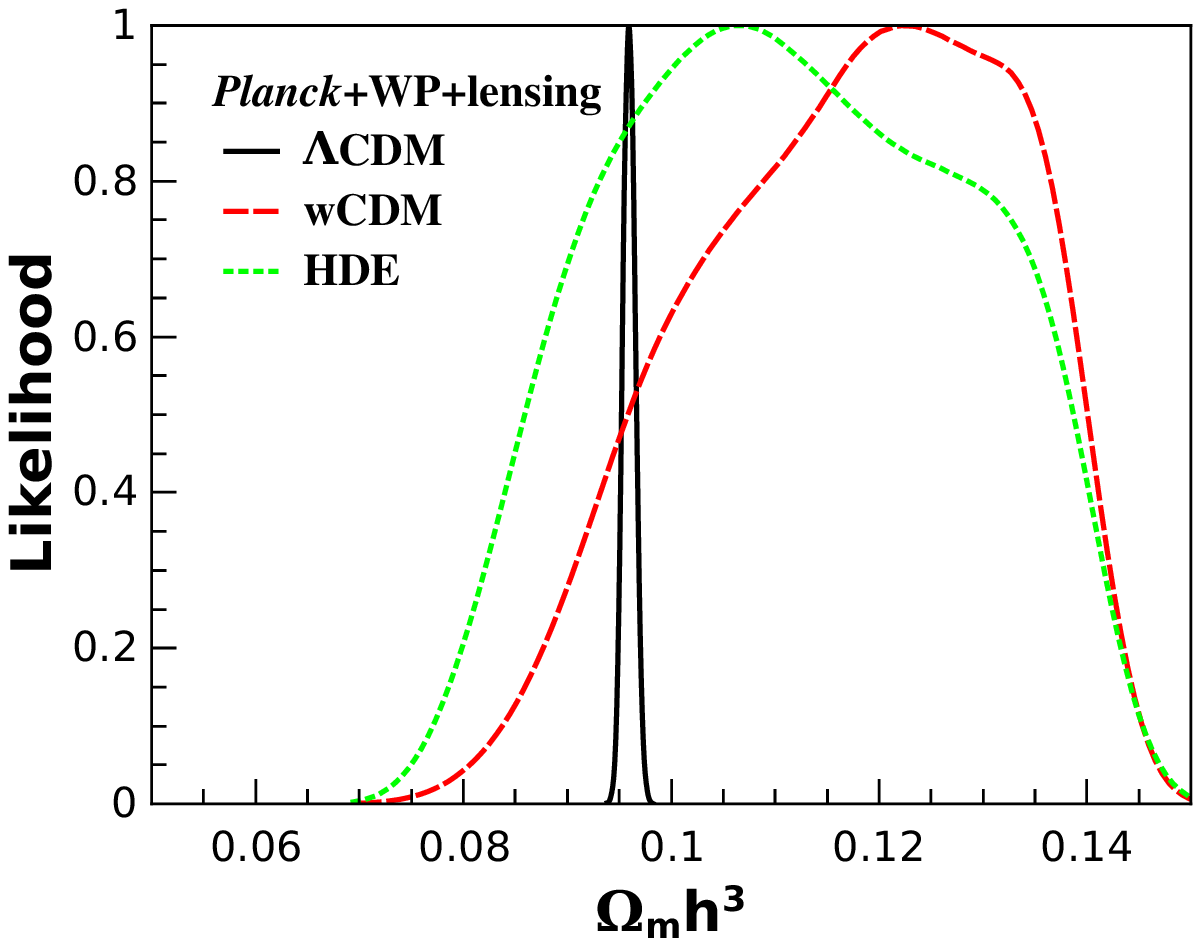}
\includegraphics[width=8cm,height=6cm]{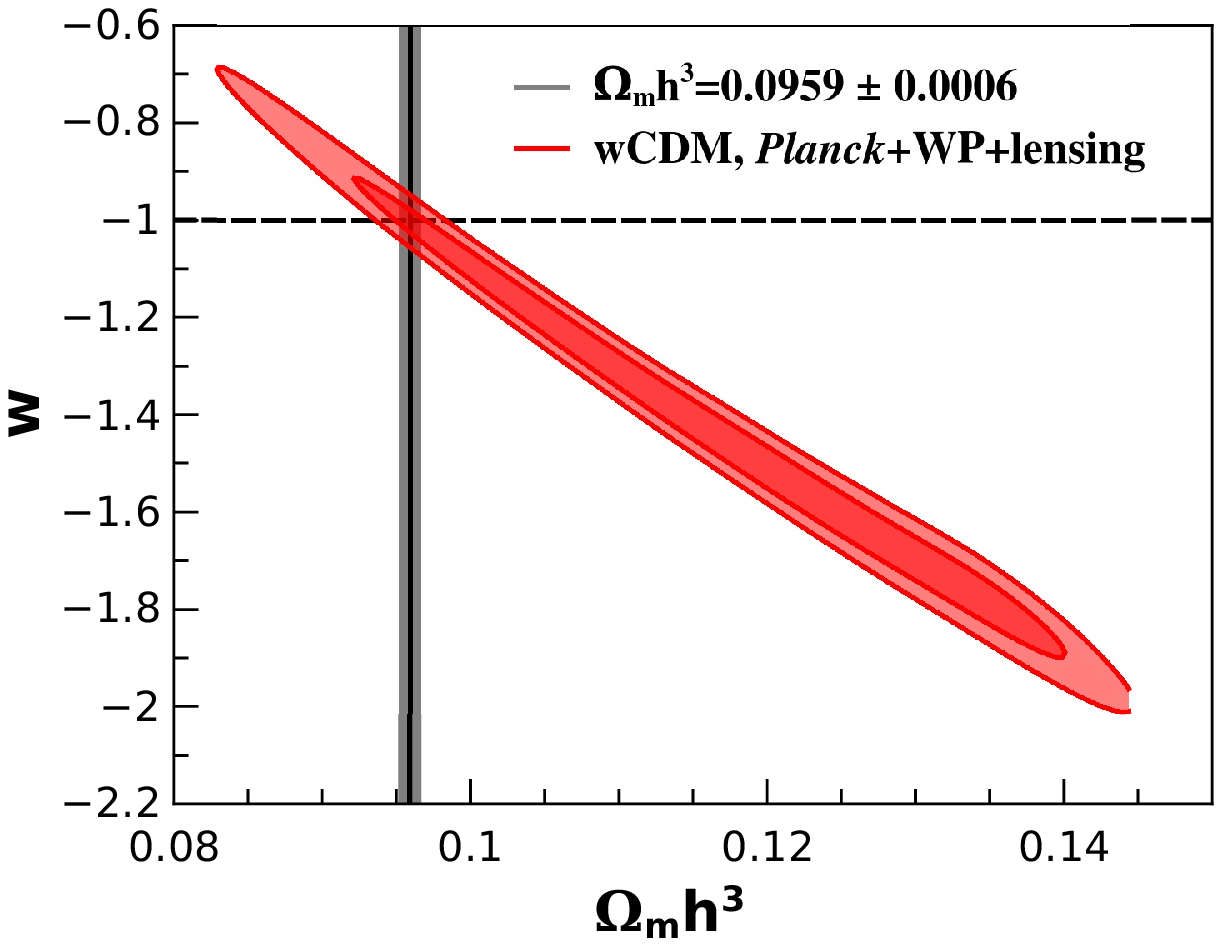}
\includegraphics[width=8cm,height=6cm]{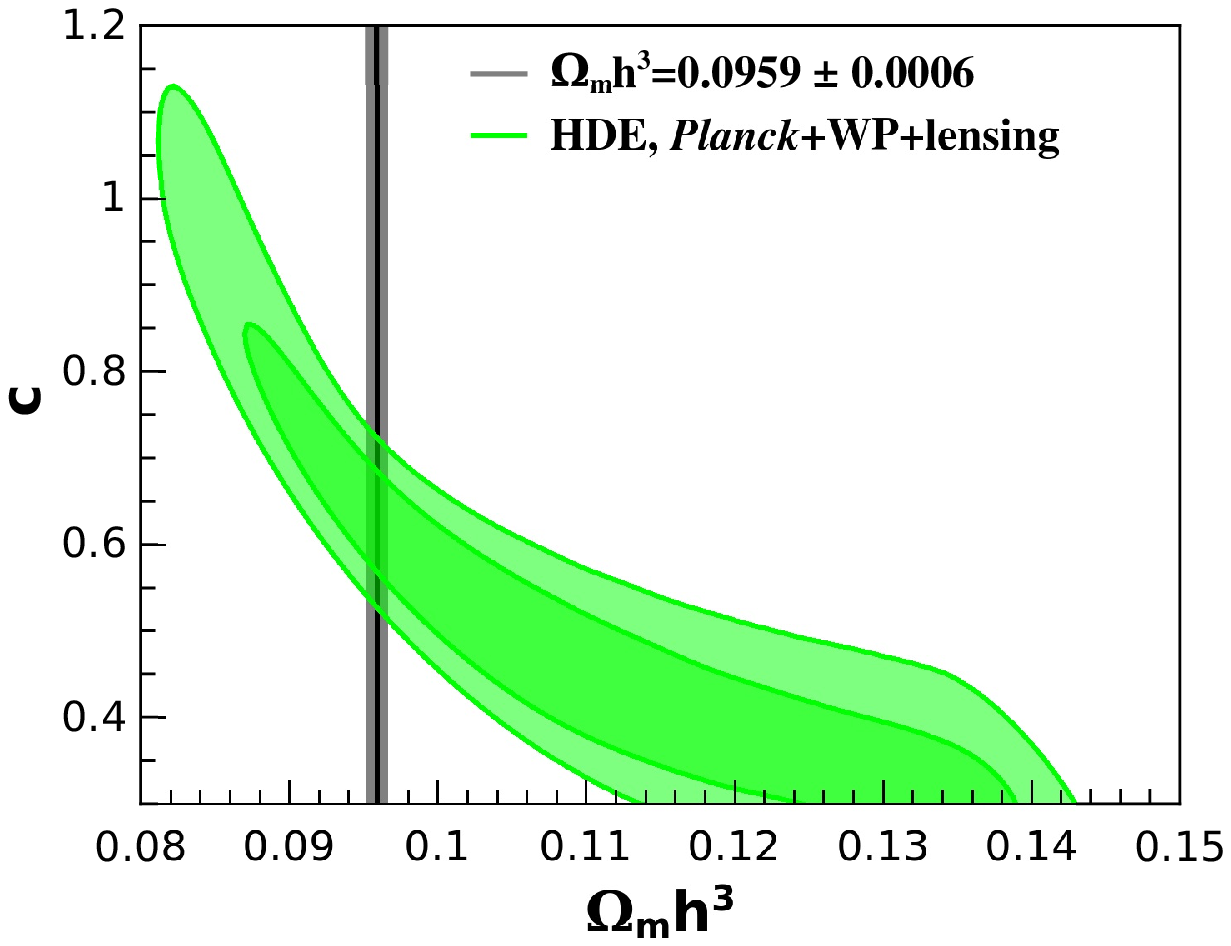}
} \caption{\label{Fig:omh23} Upper panel: Marginalized likelihood
distributions of $\Omega_{\rm m} h^2$ and $\Omega_{\rm m} h^3$,
for the $\Lambda$CDM (black solid), $w$CDM (red dashed) and HDE
(green dotted) models. Lower panel: Marginalized 68\% and 95\% CL
contours in the $\Omega_{\rm m} h^3$--$w$ and $\Omega_{\rm m}
h^3$--$c$ planes. The gray band marks the constraint $\Omega_{\rm
m} h^3=0.0959\pm0.0006$ (68\% CL; $Planck$) in the $\Lambda$CDM
model \cite{Planck16}. }
\end{figure}

Furthermore, in order to understand why the likelihood
distribution of $H_0$ is greatly widened in the $w$CDM and
HDE models,
in the upper panels of Fig.~\ref{Fig:omh23} we plot the likelihood
distributions of $\Omega_{\rm m} h^2$ and $\Omega_{\rm m} h^3$ in
the three models, constrained by {\it Planck}+WP+lensing.
Interestingly, we find similar likelihood distributions of
$\Omega_{\rm m} h^2$ in the three models, but that $\Omega_{\rm m}
h^3$ has much broader distribution in the $w$CDM and HDE models
than in the $\Lambda$CDM model. The lower panels show that the
above phenomenon is due to the strong anti-correlation between
$\Omega_{\rm m} h^3$ and dark energy parameters. In the
$\Lambda$CDM model, the precise measurement of acoustic scale in
{\it Planck} leads to a strong constraint on $\Omega_{\rm m} h^3$,
i.e., $\Omega_{\rm m} h^3=0.059\pm0.0006$ (68\% CL)
\cite{Planck16}, (shown as the gray band in the lower panels,) so
together with the constraint on $\Omega_{\rm m} h^2$ we expect a
strong constraint on $H_0$. However, when we add dark energy
parameters like $w$ or $c$ into the analysis, the strong
correlation between the parameters makes $\Omega_{\rm m} h^3$
unconstrained, and so $H_0$ also becomes unconstrained.

It is expected that the widened $H_0$ distribution is helpful in
relieving the tension between {\it Planck} and {\it {\it HST}}
observations; see \cite{PlanckAHH} for a related work. We will
discuss this topic in the next section.

\subsection{CMB lensing parameter $A_{\rm L}$}

The lensing parameter $A_{\rm L}$ is defined as a scaling parameter
of the lensing potential power spectrum \cite{DefAL},
\begin{equation}
 C^{\phi\phi}_\ell \rightarrow A_{\rm L} C^{\phi\phi}_\ell,
\end{equation}
and its theoretical expectation is $A_{\rm L}=1$.
However, by using ${\it Planck}$+WP+highL data
(``highL'' means high-$\ell$ CMB experiments; see \cite{Planck16} for details),
{\it Planck} Collaboration got $A_{\rm L}=1.23\pm0.11$ for $\Lambda$CDM,
showing a 2$\sigma$ preference for $A_{\rm L}>1$ \cite{Planck16}.
When adding the lensing measurements into the analysis, the result becomes
consistent with $A_{\rm L}=1$ at the 1$\sigma$ level; see Fig.~13 of \cite{Planck16}.

\begin{figure}[H]
\centering{
\includegraphics[height=6cm]{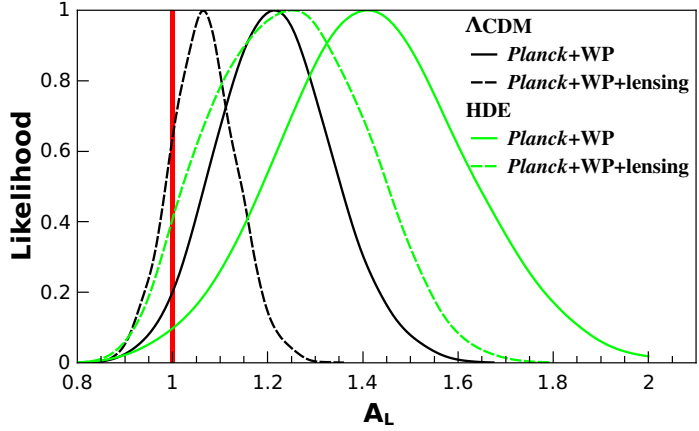}
\caption{\label{Fig:AL} Marginalized likelihood distributions of the lensing parameter $A_L$
in the $\Lambda$CDM (black) and HDE (green) models,
by using $Planck$+WP (solid) and $Planck$+WP+lensing (dashed) data.
The red solid line marks $A_L=1$.}
}
\end{figure}

To see whether HDE can help to remove this anomaly
(i.e., the preference for high $A_{\rm L}$ in the temperature power spectrum),
we repeated the similar analysis
and obtained the following results
\footnote{For convenience, in our analysis we did not use the high-$\ell$ data,
which only marginally affect the fitting results of $A_{\rm L}$.},
\begin{eqnarray}
 A_{\rm L}  &=& 1.42 \pm 0.19\ (68\%\ {\rm CL};\ {\rm HDE},\ Planck+{\rm WP}), \\
 A_{\rm L}  &=& 1.25 \pm 0.15\ (68\%\ {\rm CL};\ {\rm HDE},\ Planck+{\rm WP+lensing}).
\end{eqnarray}
As a comparison, the fitting results in $\Lambda$CDM by using the same sets of data are
\begin{eqnarray}
 A_{\rm L}  &=& 1.22 \pm 0.12\ (68\%\ {\rm CL};\ \Lambda{\rm CDM},\ Planck+{\rm WP}), \\
 A_{\rm L}  &=& 1.07 \pm 0.07\ (68\%\ {\rm CL};\ \Lambda{\rm CDM},\ Planck+{\rm WP+lensing}).
\end{eqnarray}
Corresponding marginalized posterior distributions for $A_{\rm L}$ are shown in Fig.~\ref{Fig:AL}.

Compared with $\Lambda$CDM, in the HDE model the error bars of $A_{\rm L}$ are amplified due to the extra model parameter,
but the best-fit values become even larger.
As a consequence, we find that $A_{\rm L} > 1$ at 2.2$\sigma$ and 1.7$\sigma$
by using {\it Planck}+WP and {\it Planck}+WP+lensing.
The anomaly becomes slightly worse than in $\Lambda$CDM.

\section{CMB combined with astrophysical data set
results}\label{sec:otherdata}

The CMB+Ext fitting results of the HDE model are listed in Table
\ref{Tab:CMBExt}. Best-fit values as well as 68\% CL limits for
$\Omega_{\rm m}$, $c$ and $H_0$ are listed in the columns 2--7.
The maximal likelihood values are listed in the 7$th$ column, and
the residual $\chi^2$ values, defined as $\Delta \chi^2_{\rm
CMB+Ext}\equiv\chi^2_{\rm CMB+Ext}-\chi^2_{\rm CMB}-\chi^2_{\rm
Ext}$, are listed in the last column. The results from {\it
Planck}+WP combined with external astrophysical data sets are
listed in the first nine rows, while the {\it WMAP}-9 results are
listed in the following seven rows. As a comparison, in the last
six rows we also list the results from the astrophysical data sets
only, including the fitting results of BAO, BAO+{\it HST}, SNLS3,
Union2.1, SNLS3+BAO+{\it HST}, and Union2.1+BAO+{\it HST}.

\begin{table}[!htp]
\caption{\label{Tab:CMBExt} CMB+Ext fitting results of the HDE
model. }
\begingroup
\openup 5pt
\newdimen\tblskip \tblskip=5pt
\nointerlineskip
\vskip -3mm
\scriptsize
\setbox\tablebox=\vbox{
    \newdimen\digitwidth
    \setbox0=\hbox{\rm 0}
    \digitwidth=\wd0
    \catcode`"=\active
    \def"{\kern\digitwidth}
    \newdimen\signwidth
    \setbox0=\hbox{+}
    \signwidth=\wd0
    \catcode`!=\active
    \def!{\kern\signwidth}
\halign{ \hbox to
2.3in{$#$\leaderfil}\tabskip=1.5em&\hfil$#$\hfil&\hfil$#$\hfil&\hfil$#$\hfil&\hfil$#$\hfil&\hfil$#$\hfil&\hfil$#$\hfil\tabskip=0pt&\hfil$#$\hfil\tabskip=0pt&\hfil$#$\hfil\tabskip=0pt\cr
\noalign{\doubleline} \multispan1\hfil \hfil&\multispan2\hfil
$\Omega_{\rm m}$ \hfil&\multispan2\hfil $c$ \hfil&\multispan2\hfil
$H_0$ \hfil\cr \noalign{\vskip -3pt}
\omit&\multispan2\hrulefill&\multispan2\hrulefill&\multispan2\hrulefill&\omit&\omit\cr
\omit\hfil Data\hfil&\omit\hfil Best fit\hfil&\omit\hfil 68\%
limits\hfil&\omit\hfil Best fit\hfil&\omit\hfil 68\%
limits\hfil&\omit\hfil Best fit\hfil&\omit\hfil 68\% limits\hfil
&\omit\hfil $\ \ \ \ \ \ -\ln \mathcal{L}_{max}\ \ \ \ \
$\hfil&\omit\hfil $\Delta \chi^2$ $\ ^a$ \hfil\cr \noalign{\vskip
5pt\hrule\vskip 3pt}
Planck{\rm +WP+BAO} & 0.270 & 0.254\pm0.024 & 0.506 & 0.484\pm0.070 & 72.63 & 75.06\pm3.82 & 4903.6 & 1.7 \cr
Planck{\rm +WP+BAO+lensing} & 0.262 & 0.256\pm0.022 & 0.498 & 0.494\pm0.062 & 73.62 & 74.65\pm3.39 & 4908.4 & 1.9 \cr
Planck{\rm +WP+{\it {\it HST}}} & 0.266 & 0.257\pm0.019 & 0.463 & 0.474\pm0.049 & 73.78 & 74.77\pm2.68 & 4902.8 & 0.3\ ^c \cr
Planck{\rm +WP+{\it {\it HST}}+lensing} & 0.260 & 0.256\pm0.019 & 0.498 & 0.489\pm0.048 & 73.81 & 74.62\pm2.69 & 4907.9 & 1.1\ ^c \cr
Planck{\rm +WP+BAO+{\it {\it HST}}} & 0.252 & 0.255\pm0.014 & 0.470 & 0.481\pm0.046 & 75.22 & 74.75\pm2.19 & 4903.6 & 0.9 \cr
Planck{\rm +WP+BAO+{\it {\it HST}}+lensing} & 0.245 & 0.255\pm0.013 & 0.481 & 0.495\pm0.039 & 75.83 & 74.5\pm2.0 & 4908.5 & 1.3 \cr
Planck{\rm +WP+SNLS3} & 0.300 & 0.305\pm0.019 & 0.584 & 0.594\pm0.051 & 68.81 & 68.46\pm1.93 & 5115.8 & 6.4 \cr
Planck{\rm +WP+SNLS3+lensing} & 0.310 & 0.301\pm0.019 & 0.610 & 0.603\pm0.049 & 67.73 & 68.66\pm1.92 & 5120.8 & 7.3 \cr
Planck{\rm +WP+Union2.1} & 0.327 & 0.324\pm0.021 & 0.618 & 0.642\pm0.066 & 66.35 & 66.74\pm1.94 & 5176.1 & 1.6 \cr
Planck{\rm +WP+Union2.1+lensing} & 0.321 & 0.321\pm0.021 & 0.617 & 0.645\pm0.063 & 66.72 & 66.68\pm2.03 & 5181.6 & 3.4 \cr
Planck{\rm +WP+SNLS3+BAO+{\it {\it HST}}+lensing} & 0.269 & 0.275\pm0.011 & 0.583 & 0.563\pm0.035 & 72.41 & 71.46\pm1.37 & 5123.2 & 10.9\ ^d \cr
Planck{\rm +WP+Union2.1+BAO+{\it {\it HST}}+lensing} & 0.276 & 0.281\pm0.012 & 0.551 & 0.577\pm0.039 & 71.49 & 70.68\pm1.40 & 5185.3 & 9.6\ ^d \cr
\noalign{\vskip 5pt\hrule\vskip 3pt}
{\rm {\it WMAP}\textendash9+BAO} & 0.274 & 0.284\pm0.021 & 0.623 &
0.746\pm0.165 & 70.41 & 68.93\pm3.18 & 3779.6 & 0.9\cr {\rm {\it
WMAP}\textendash9+{\it {\it HST}}} & 0.251 & 0.250\pm0.020 & 0.552
& 0.569\pm0.086 & 73.98 & 73.99\pm2.71 & 3779.1 & 0.2\ ^c \cr {\rm
{\it WMAP}\textendash9+BAO+{\it {\it HST}}} & 0.255 &
0.259\pm0.015 & 0.534 & 0.567\pm0.081 & 73.65 & 72.96\pm2.37 &
3779.7 & 0.3\ ^c \cr {\rm {\it WMAP}\textendash9+SNLS3} & 0.277 &
0.280\pm0.022 & 0.664 & 0.696\pm0.078 & 69.79 & 69.44\pm2.23 &
3990.6 & 3.5\cr {\rm {\it WMAP}\textendash9+Union2.1} & 0.304 &
0.299\pm0.023 & 0.767 & 0.782\pm0.105 & 66.76 & 67.24\pm2.18 &
4051.6 & 0.1 \cr {\rm {\it WMAP}\textendash9+SNLS3+BAO+{\it {\it
HST}}} & 0.269 & 0.270\pm0.011 & 0.626 & 0.645\pm0.060 & 70.89 &
70.89\pm1.46 & 3992.2 & 5.6\ ^d \cr {\rm {\it
WMAP}\textendash9+Union2.1+BAO+{\it {\it HST}}} & 0.276 &
0.276\pm0.011 & 0.659 & 0.711\pm0.074 & 70.17 & 69.64\pm1.37 &
4054.3 & 4.3\ ^d \cr \noalign{\vskip 3pt\hrule\vskip 5pt} {\rm
BAO} & 0.227 & 0.215\pm0.124 & 2.391 & 1.579\pm0.772 & \times ^e
 &
\times & 0.1 & --\cr {\rm BAO+{\it {\it HST}}} & 0.289 &
0.332\pm0.974 & 0.552 & 0.666\pm0.241 & 73.56 & 73.49\pm2.38 & 0.6
& --\cr {\rm SNLS3} & 0.129 & 0.118\pm0.072 & 1.294 &
1.519\pm0.514 & \times & \times & 209.8 & --\cr {\rm Union2.1} &
0.256 & 0.173\pm 0.099 & 0.024 & 1.68\pm0.78 & \times & \times &
272.5 & --\cr {\rm SNLS3+BAO+{\it {\it HST}}} & 0.294 &
0.295\pm0.029 & 0.612 & 0.622\pm0.071 & 72.27 & 72.37\pm2.36 &
212.5 & 4.1\ ^b \cr {\rm Union2.1+BAO+{\it {\it HST}}} & 0.323 &
0.326\pm0.030 & 0.608 & 0.633\pm0.086 & 73.42 & 73.09\pm2.36 &
273.6 & 1.0\ ^b \cr \noalign{\vskip 5pt\hrule\vskip 3pt}
} 
} 
\endPlancktable
\endgroup
\leftline{\noindent$\ ^a$ $\Delta \chi^2_{\rm CMB+Ext}\equiv\chi^2_{\rm CMB+Ext}-\chi^2_{\rm CMB}-\chi^2_{\rm Ext}$.}
\leftline{\noindent$\ ^b$ $\Delta \chi^2_{\rm SNIa+BAO+{\it HST}}\equiv\chi^2_{\rm SNIa+BAO+{\it HST}}-\chi^2_{\rm SNIa}-\chi^2_{\rm BAO+{\it HST}}$.}
\leftline{\noindent$\ ^c$ $\Delta \chi^2_{\rm CMB+{\it HST}}\equiv\chi^2_{\rm CMB+{\it HST}}-\chi^2_{\rm CMB}$.}
\leftline{\noindent$\ ^d$ $\Delta \chi^2_{\rm CMB+SNIa+BAO+{\it HST}}\equiv\chi^2_{\rm CMB+SNIa+BAO+{\it HST}}-\chi^2_{\rm CMB}-\chi^2_{\rm SNIa}-\chi^2_{\rm BAO+{\it HST}}$.}

\leftline{\noindent$\ ^e$ The cross ``$\times$'' indicates that
the parameter is unconstrained by the chosen data sets.}

\end{table}

\begin{figure}[H]
\centering{
\includegraphics[width=8cm,height=6cm]{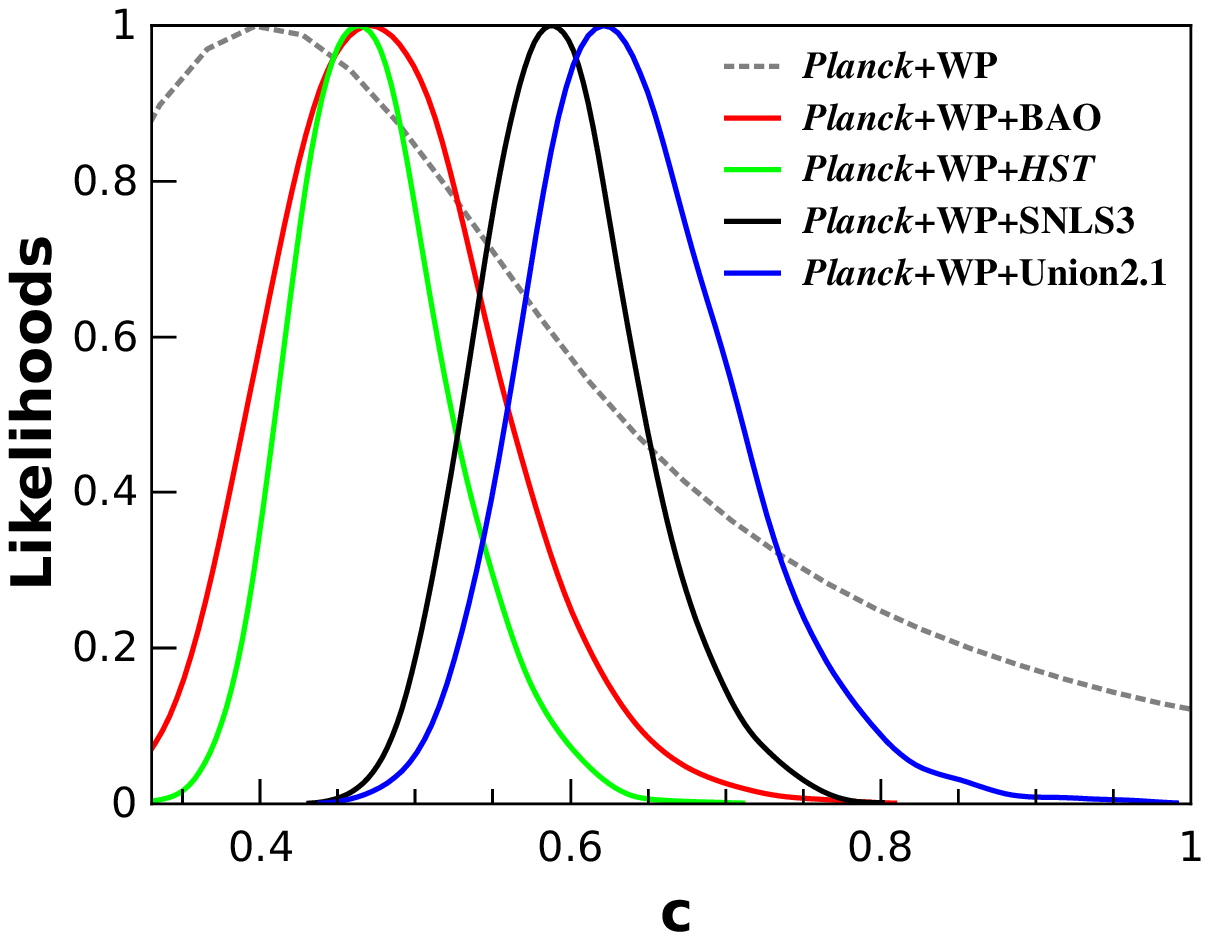}
\includegraphics[width=8cm,height=6cm]{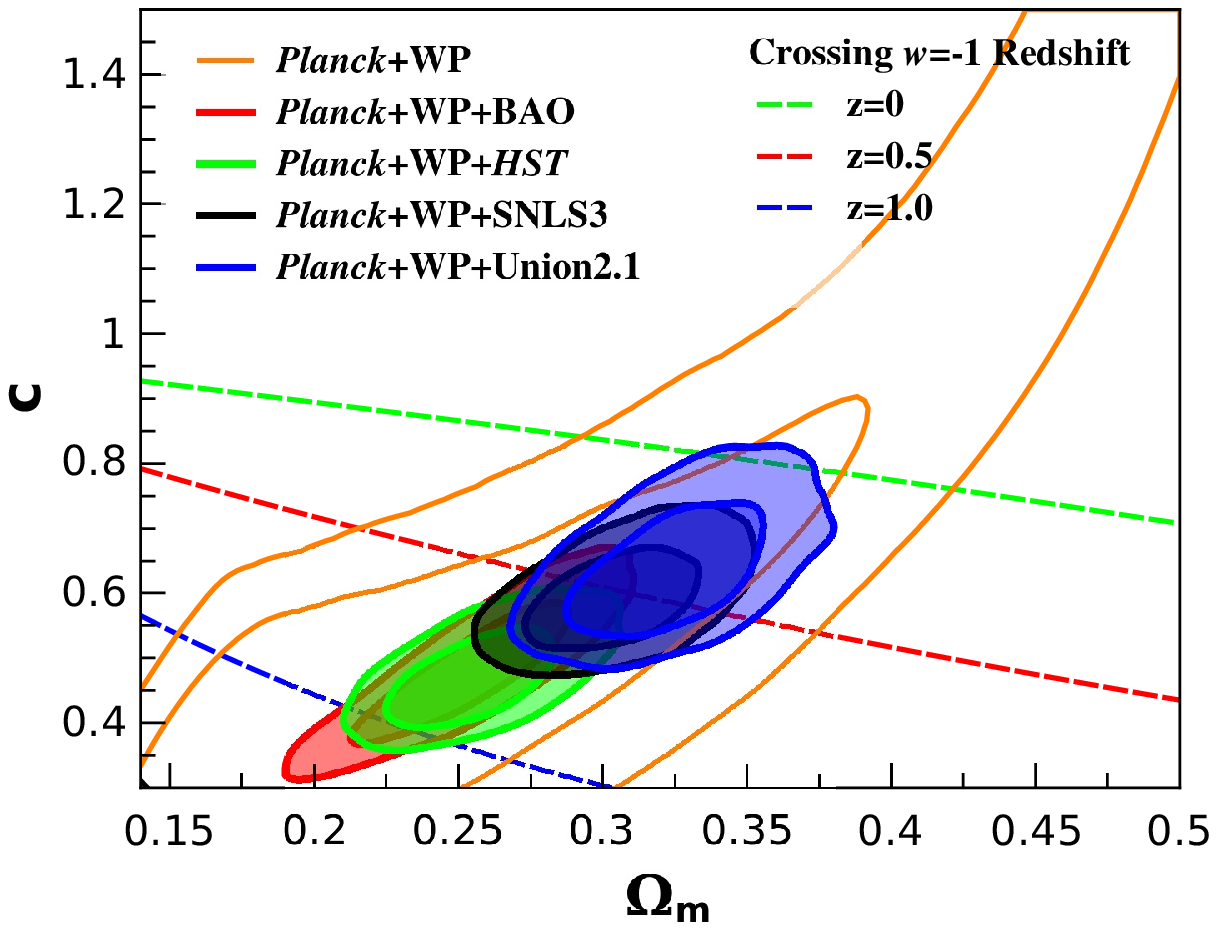}}
\caption{\label{Fig:cmbcombined} Fitting results of the HDE model,
from {\it Planck}+WP combined with external astrophysical data
sets of BAO (red), {\it HST} (green), SNLS3 (black) and Union2.1
(blue).
Left panel: Marginalized likelihood distributions of $c$. Right
panel: Marginalized 68\% and 95\% CL contours in the $\Omega_{\rm
m}$--$c$ plane.}
\end{figure}

We find that adding external astrophysical datset 
reduces the error of $c$ to 0.05--0.07. The likelihood
distributions of $c$ and the $\Omega_{\rm m}$--$c$ contours for
{\it Planck}+WP and {\it Planck}+WP+Ext are plotted in
Fig.~\ref{Fig:cmbcombined}. The best-fit values of $c$ for {\it
Planck}+WP+BAO and {\it Planck}+WP+{\it HST} constraints are
around 0.5, while the values for {\it Planck}+WP+SNLS3 and {\it
Planck}+WP+Union2.1 constraints are around 0.6. As a comparison,
the best-fit value of $c$ from {\it WMAP}-9 combined one Ext is
larger, i.e., 0.55--0.77, and the error is also larger, i.e.,
0.08--0.17.

We find that the {\it Planck} lensing data are helpful in improving
the constraint on $c$. By adding the lensing data into the analysis
of {\it Planck}+WP combined with the Ext, such as BAO, {\it HST},
BAO+{\it HST}, SNLS3 and Union2.1, the constraint results are
improved by 11\%, 2\%, 15\%, 4\% and 5\%, respectively.

\begin{figure}[H]
\centering{
\includegraphics[width=8cm,height=6cm]{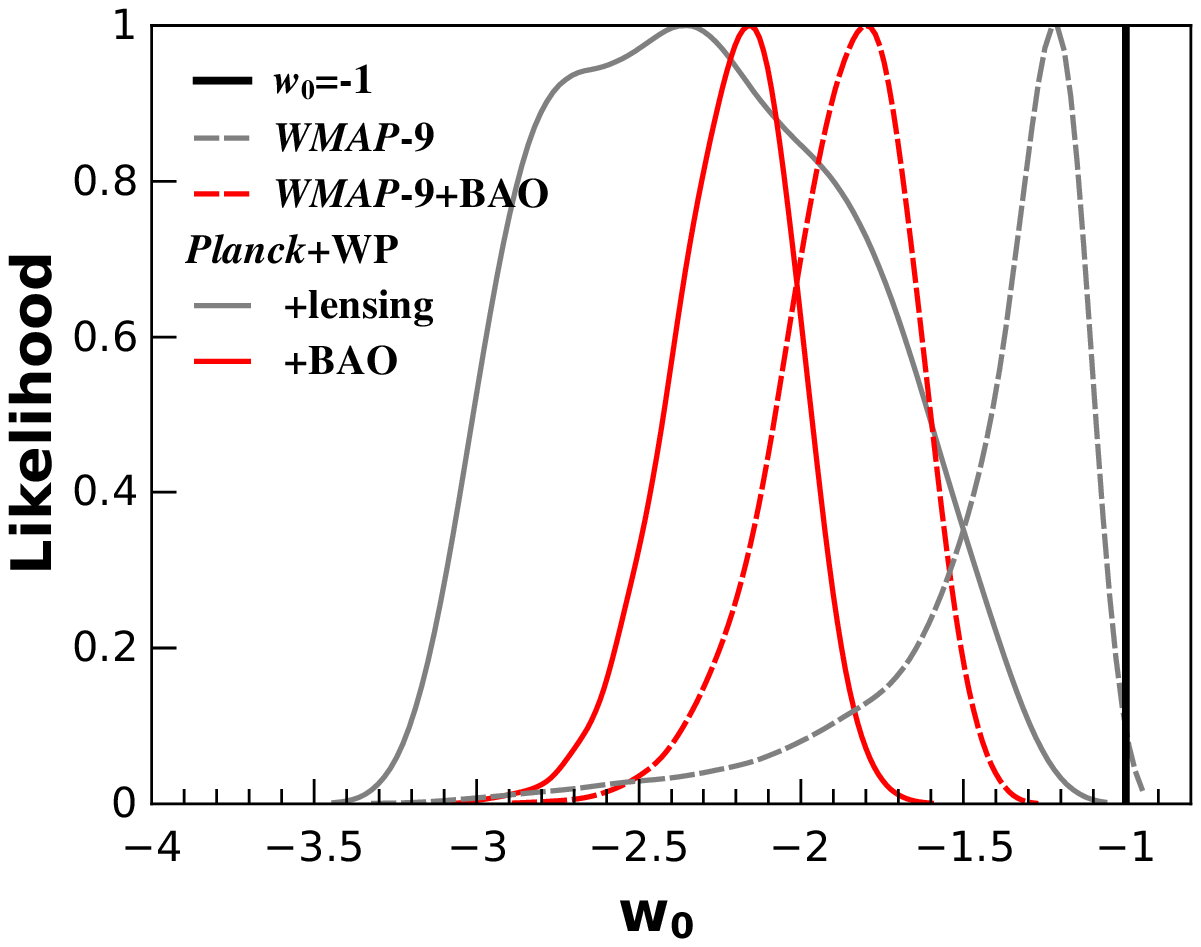}
\includegraphics[width=8cm,height=6cm]{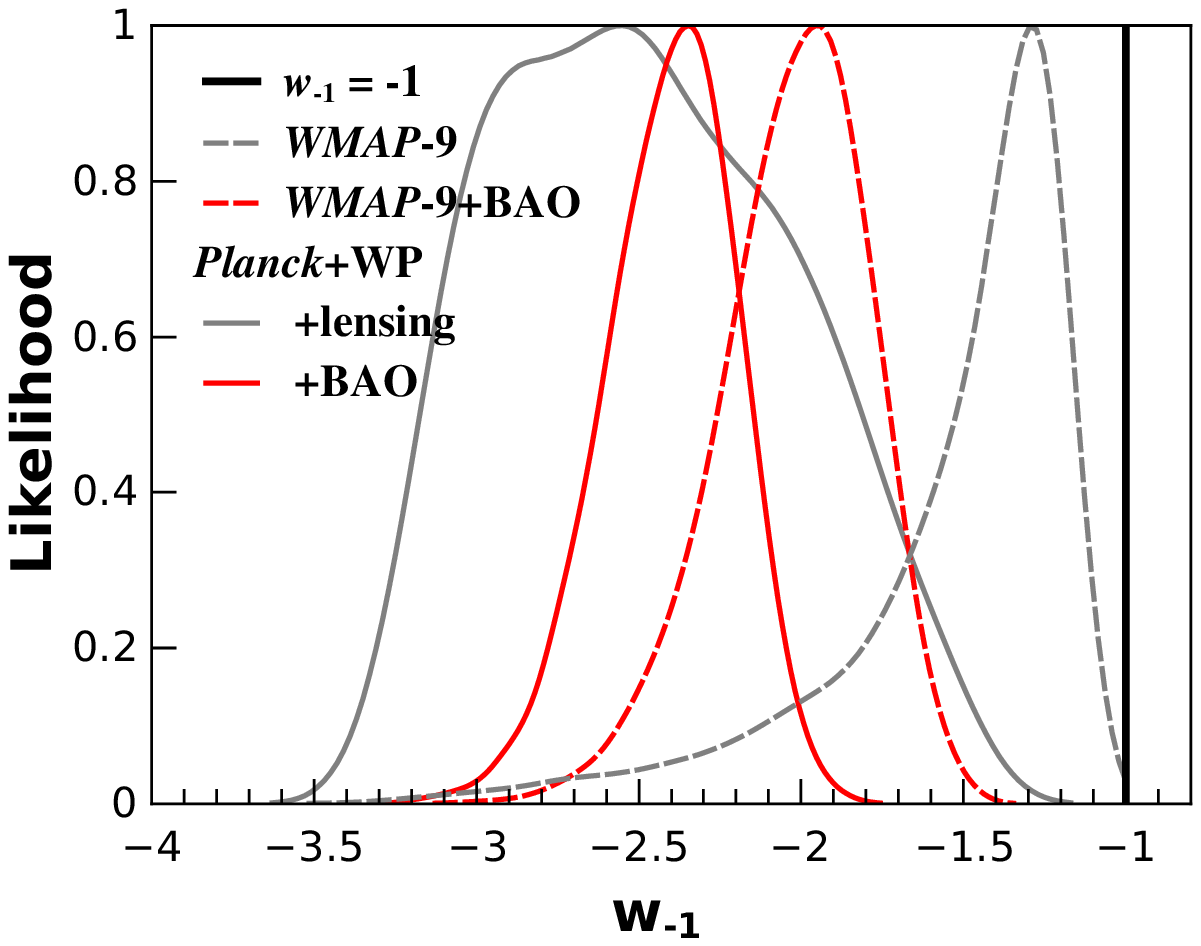}
\includegraphics[width=15cm]{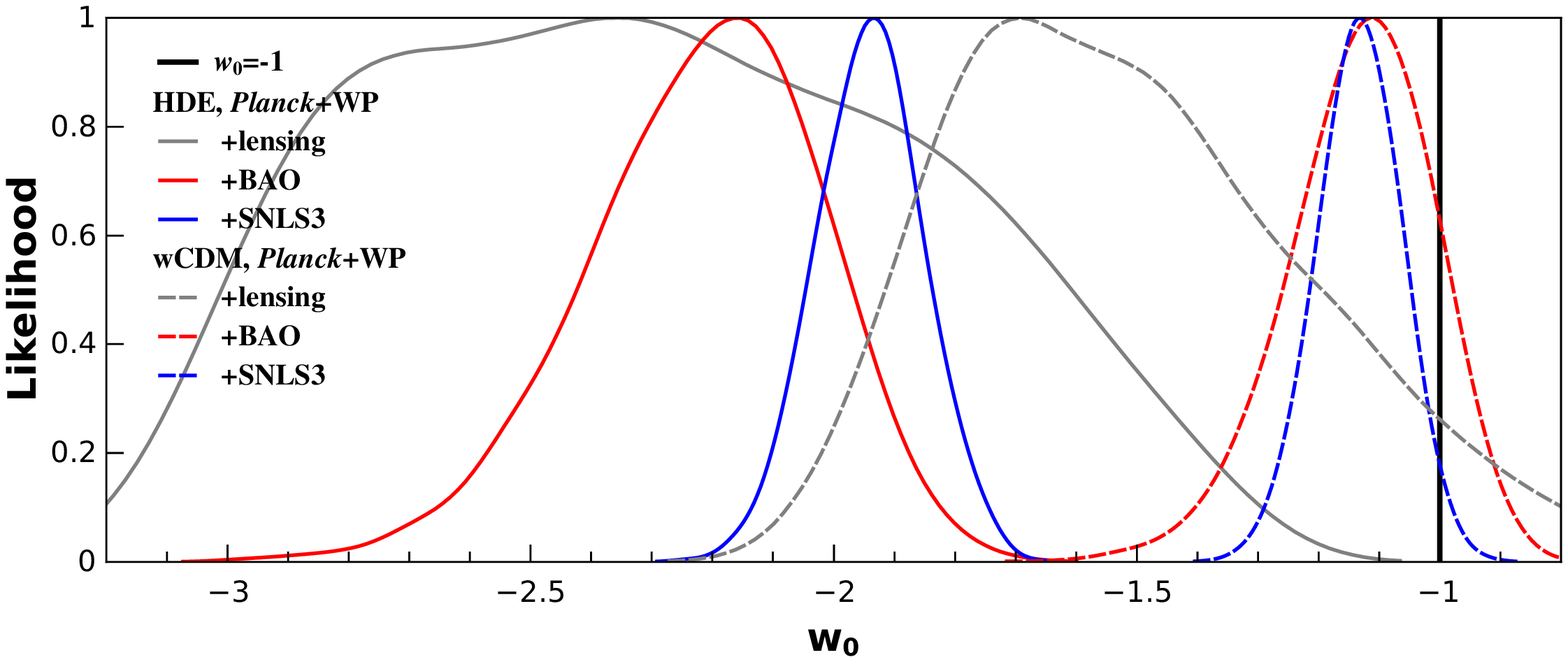}}
\caption{\label{Fig:wcmbcombined} Marginalized likelihood
distributions of dark energy equation of state at $z=0$ (upper left
and lower panels) and $z=-1$ (upper right panel). In the two upper
panels, the results of {\it Planck}+WP combined with lensing (gray
solid) and BAO (red solid) are shown. In the lower panel, we also
plot the {\it Planck}+WP+SNLS3 results (blue). The black thick line
marks $w=-1$. As comparisons, the {\it WMAP}-9 (gray dashed) and
{\it WMAP}-9+BAO (red dashed) results are plotted in the two upper panels,
and the $w$CDM results (dashed lines) are plotted in the lower panel. }
\end{figure}

To see the dynamical behavior of HDE, in Fig.~\ref{Fig:wcmbcombined}
we plot the likelihood distributions of the dark energy equation of
state at $z=0$ (upper left and lower panels) and $z=-1$ (upper right
panel). We find that, by using the {\it Planck} data, a phantom-like
holographic dark energy is favored at high confidence level in both
current and future epochs. The result of $w_0<-1$ can be obtained at
more than 2$\sigma$ level by using {\it Planck}+WP+lensing, even
without any Ext combined. These are different from that of the
$w$CDM results (dashed lines in the lower panel), where $w=-1$ is
still consistent with the fitting results at a relatively high
confidence level.

Furthermore, to investigate the tension between CMB and Ext, in the
last column of Table~\ref{Tab:CMBExt} we list the $\Delta \chi^2$
values for the different combinations. In most combinations we find
a small $\Delta \chi^2$, except for the CMB+SNLS3 results, where
$\chi^2_{\rm CMB+SNLS3}-\chi^2_{\rm CMB}-\chi^2_{\rm SNLS3}=6.4$,
7.3, and 3.5 for {\it Planck}+WP, {\it Planck}+WP+lensing and {\it WMAP}-9,
implying an evident tension.
For no-CMB constraints, the result $\chi^2_{\rm
SNLS3+BAO+{\it HST}}-\chi^2_{\rm SNLS3}-\chi^2_{\rm BAO+{\it
HST}}=4.1$ means that SNLS3 is also in tension with BAO+{\it HST}.
The {\it HST} combined results lead to $\chi^2_{\rm CMB+{\it
HST}}-\chi^2_{\rm CMB}=1.7$, 1.1 and 0.2 for {\it Planck}+WP,
{\it Planck}+WP+lensing and {\it WMAP}-9, implying that there is no severe tension between {\it HST}
and CMB in the HDE model.

In the following, we will discuss the fitting results in detail. We
will discuss the fitting results of CMB combined with BAO and {\it
HST} in the first subsection, and the fitting results of CMB
combined with SNLS3 and Union2.1 in the second subsection.

\subsection{Combined with BAO and {\it HST}}


\begin{figure}[H]
\centering{
\includegraphics[width=8cm,height=6cm]{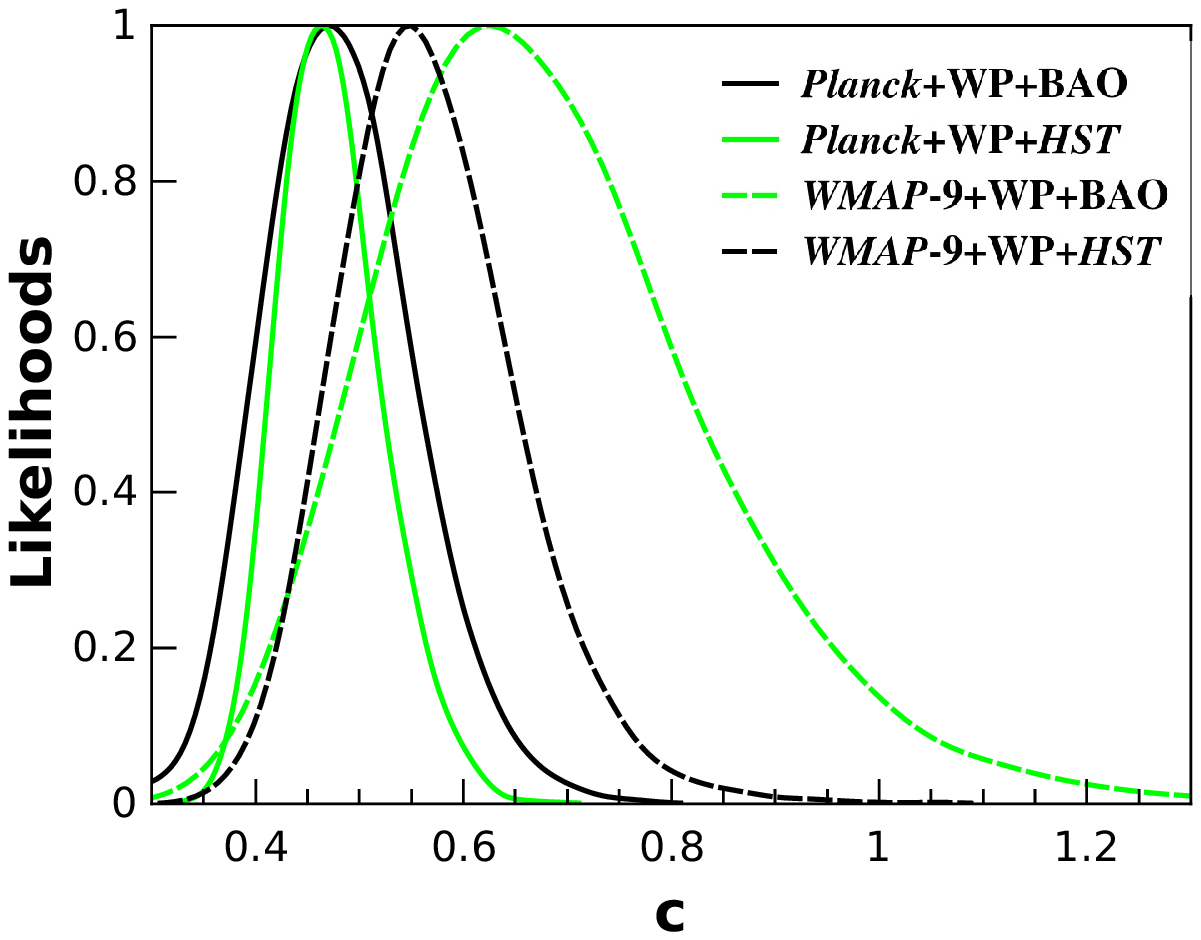}
\includegraphics[width=8cm,height=6cm]{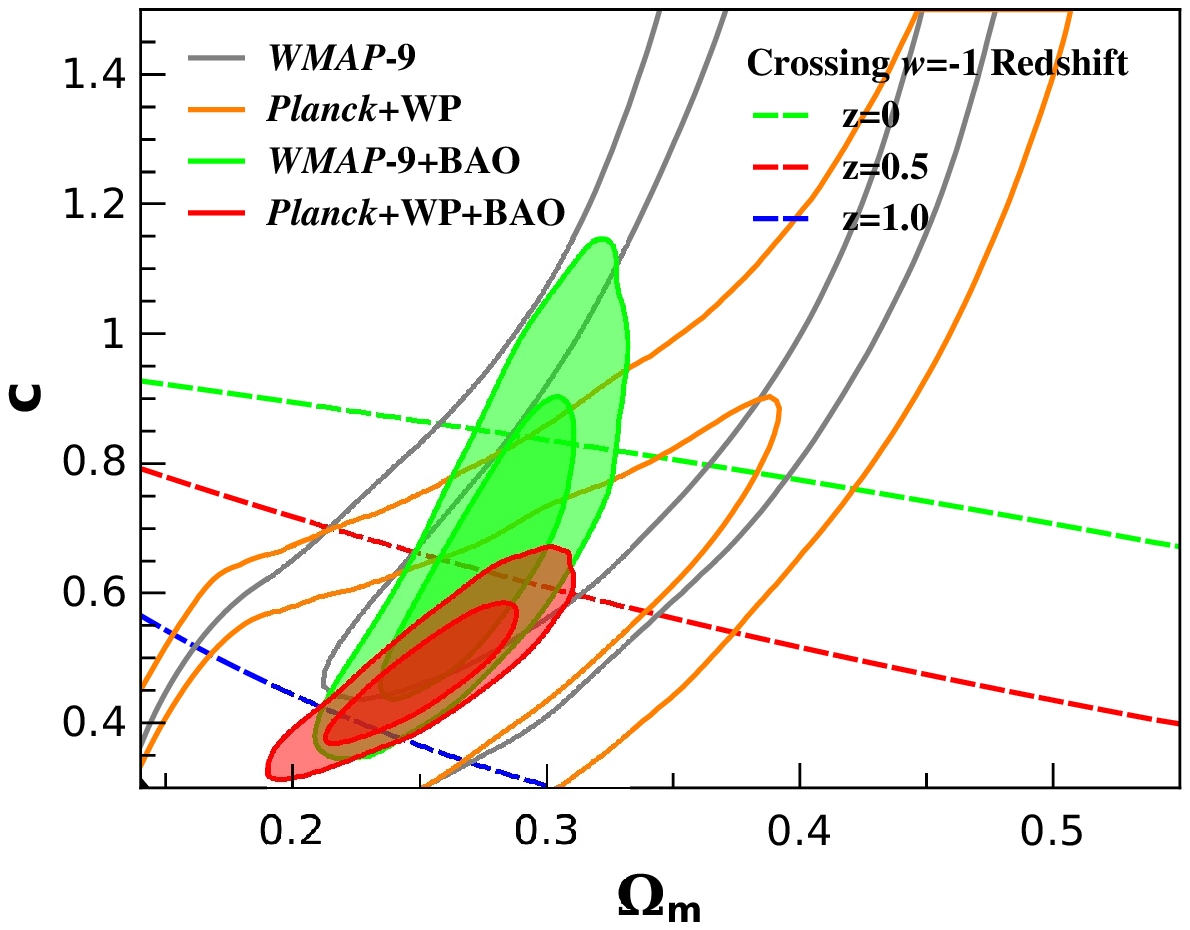}
\includegraphics[width=8cm,height=6cm]{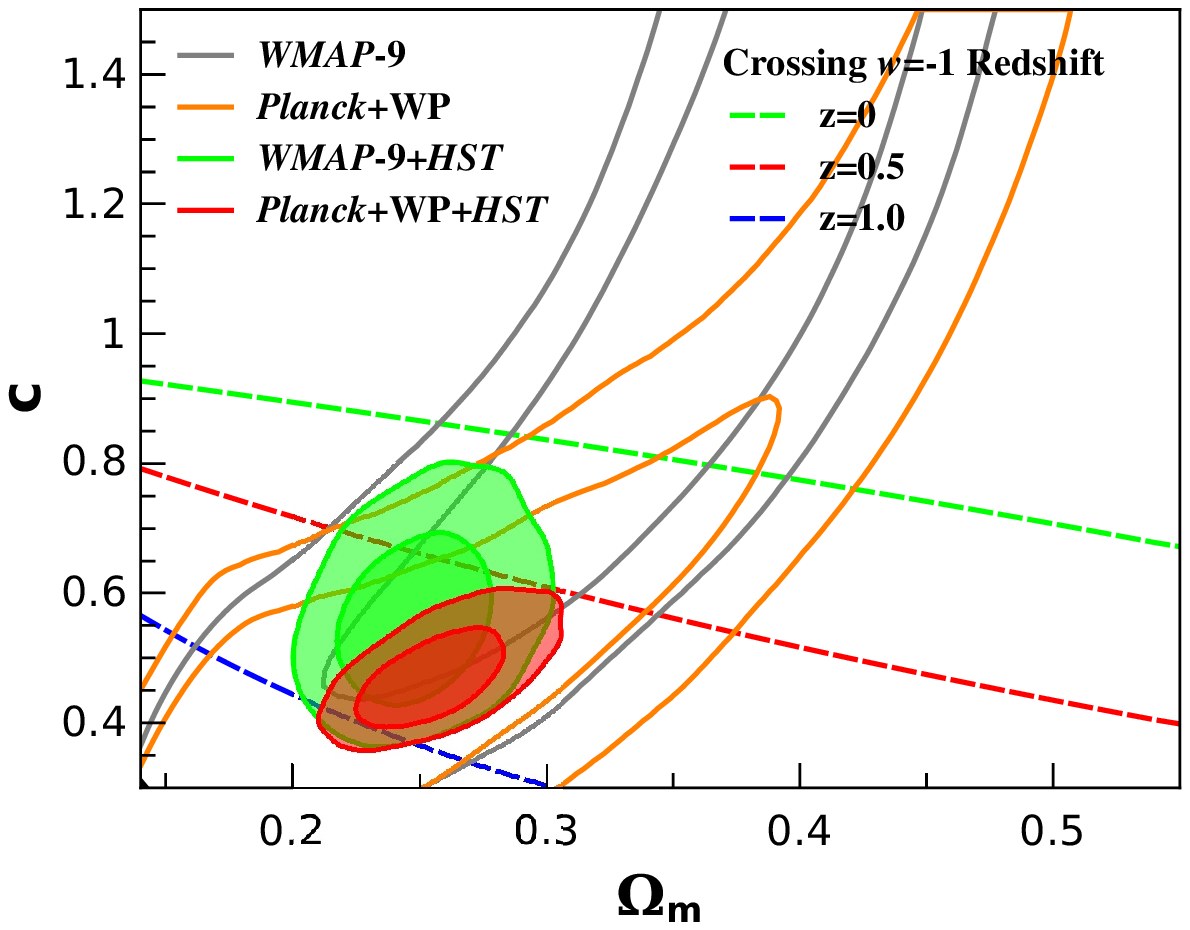}
\includegraphics[width=8cm,height=6cm]{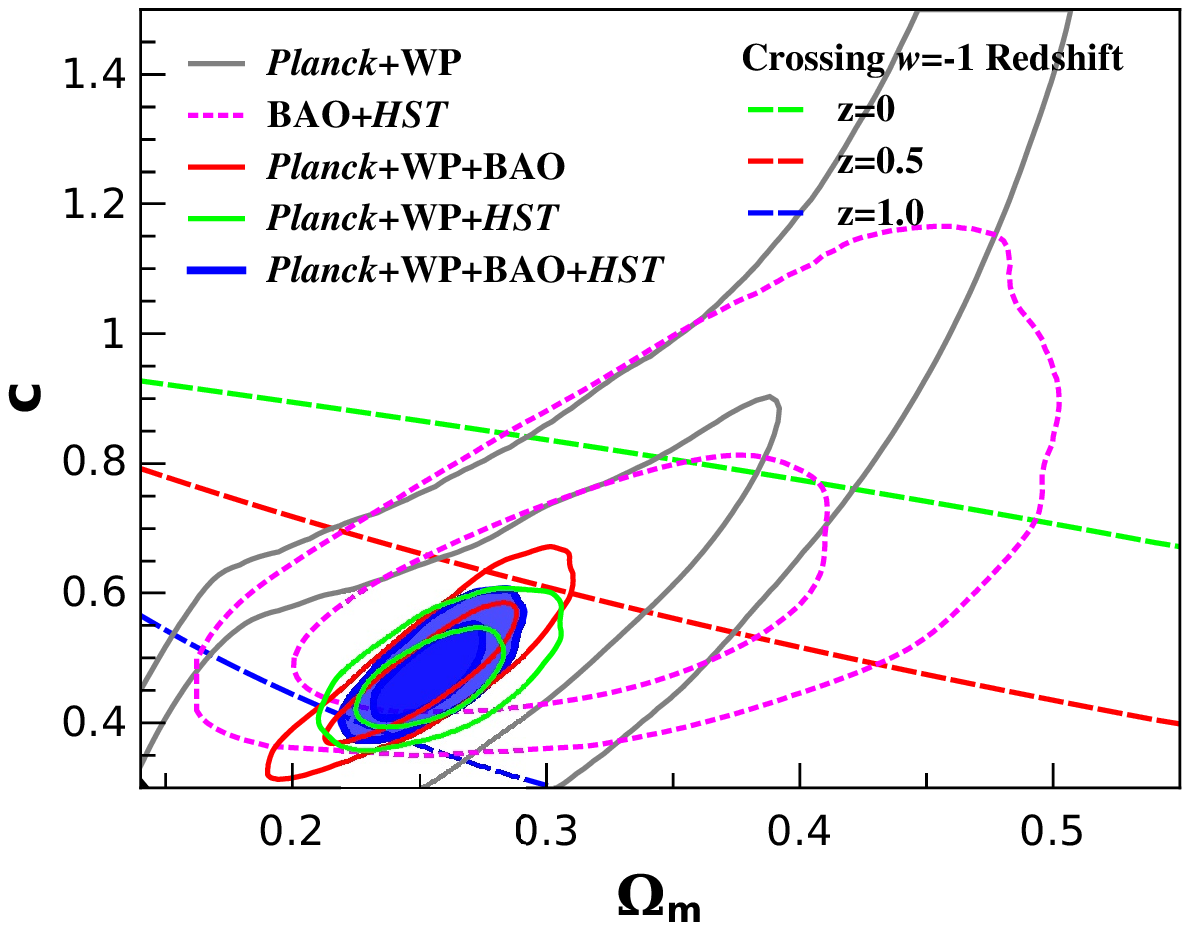}}
\caption{\label{Fig:cmbbaohst} Fitting results of the HDE model,
from CMB combined with BAO and {\it HST}. The upper-left panel
shows the marginalized distributions of $c$. The other three
panels show the marginalized 68\% and 95\% CL contours in the
$\Omega_{\rm m}$--$c$ plane, including the CMB+BAO results for
{\it Planck} and {\it WMAP}-9 (upper-right), CMB+{\it HST} results
for {\it Planck} and {\it WMAP}-9 (lower-left), and the CMB+{\it
HST}/BAO results for {\it Planck} (lower-right).
}
\end{figure}

At the 68\% CL, we obtain $c=0.484\pm0.070$ ({\it Planck}+WP+BAO),
$c=0.474\pm0.049$ ({\it Planck}+WP+{\it HST}), $c=0.746\pm0.165$
({\it WMAP}-9+BAO) and $c=0.569\pm0.086$ ({\it WMAP}-9+{\it HST}).
Compared with the {\it WMAP}-9 results, the best-fit values of $c$
from the {\it Planck} data are smaller by 0.1--0.3, and the error
bars are reduced by 40\%--60\%. These results can be seen clearly
in the likelihood distributions plotted in the upper-left panel of
Fig.~\ref{Fig:cmbbaohst}.

The other three panels of Fig.~\ref{Fig:cmbbaohst} show the 68\%
and 95\% CL contours in the $\Omega_{\rm m}$--$c$ plane, including
the CMB+BAO results for {\it Planck} and {\it WMAP}-9
(upper-right), CMB+{\it HST} results for {\it Planck} and {\it
WMAP}-9 (lower-left), and the {\it Planck}+{\it HST}/BAO results
(lower-right). Interestingly, in the lower-right panel we see that
the {\it Planck}+WP+BAO (red solid) and {\it Planck}+WP+{\it HST}
(green solid) contours lie in the same position, showing that {\it
Planck}+WP+BAO and {\it Planck}+WP+{\it HST} lead to consistent
fitting results.
This figure also shows a consistent overlap of {\it Planck}+WP (gray
solid) and BAO+{\it HST} (purple dotted). The combined {\it
Planck}+WP+BAO+{\it HST} (blue filled region) data lead to a self-consistent constraint,
$c=0.481\pm0.046$.

Moreover, we can further tighten the constraints by adding the
lensing data into the the analysis. Table~\ref{Tab:CMBExt} shows
that, by adding the lensing data, the {\it Planck}+WP+BAO
constraint on $c$ is improved from $0.484\pm0.070$ to
$0.494\pm0.062$, and the {\it Planck}+WP+BAO+{\it HST} result is
improved from $0.481\pm0.046$ to $c=0.495\pm0.039$. The error bars
are reduced by 12\%--15\%. The $\Delta \chi^2$ values for the two
lensing combined results are 1.9 and 1.3, respectively, showing a
good consistency. Actually, the constraint result,
$c=0.495\pm0.039$, from {\it Planck}+WP+BAO+{\it HST}+lensing is
our tightest {\it self-consistent} constraint on $c$. If we
further add the supernova data set into the analysis, the error
bars can be slightly reduced, but a significant inconsistency
among the data sets appears. This result also has 35\%--50\%
smaller error bars compared with the {\it WMAP}-9 all-combined
results, where the constraints on $c$ are $c=0.645\pm0.060$ and
$c=0.711\pm0.074$ for {\it WMAP}-9+BAO+{\it HST} combined with
SNLS3 and Union2.1, respectively.


\begin{figure}[H]
\centering{
\includegraphics[width=16cm]{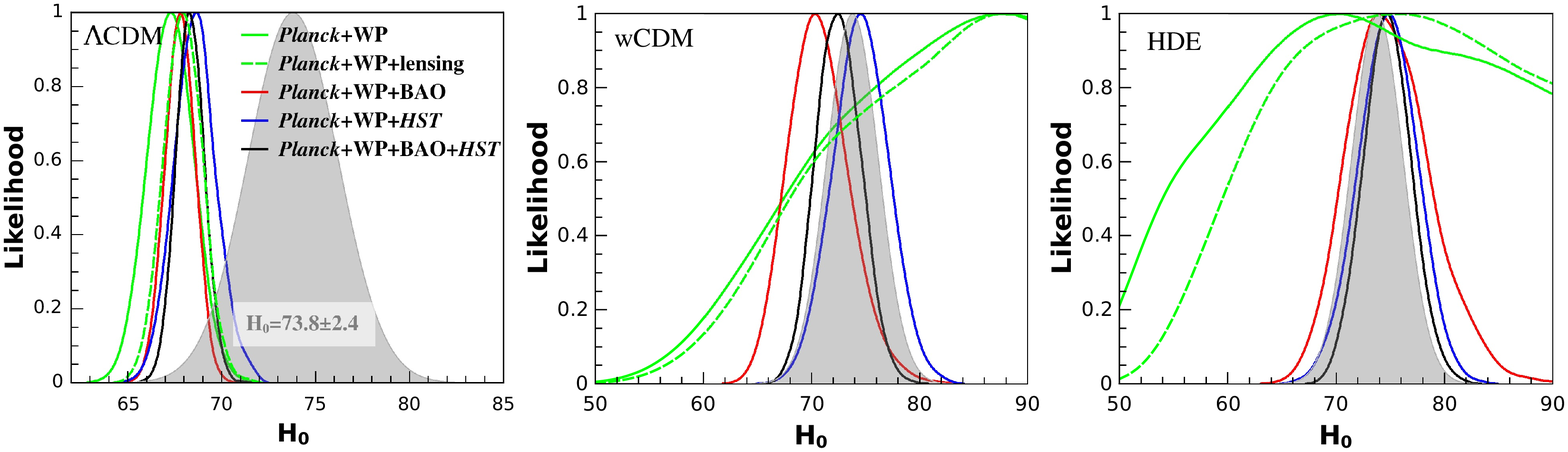}}
\includegraphics[width=8cm,height=6cm]{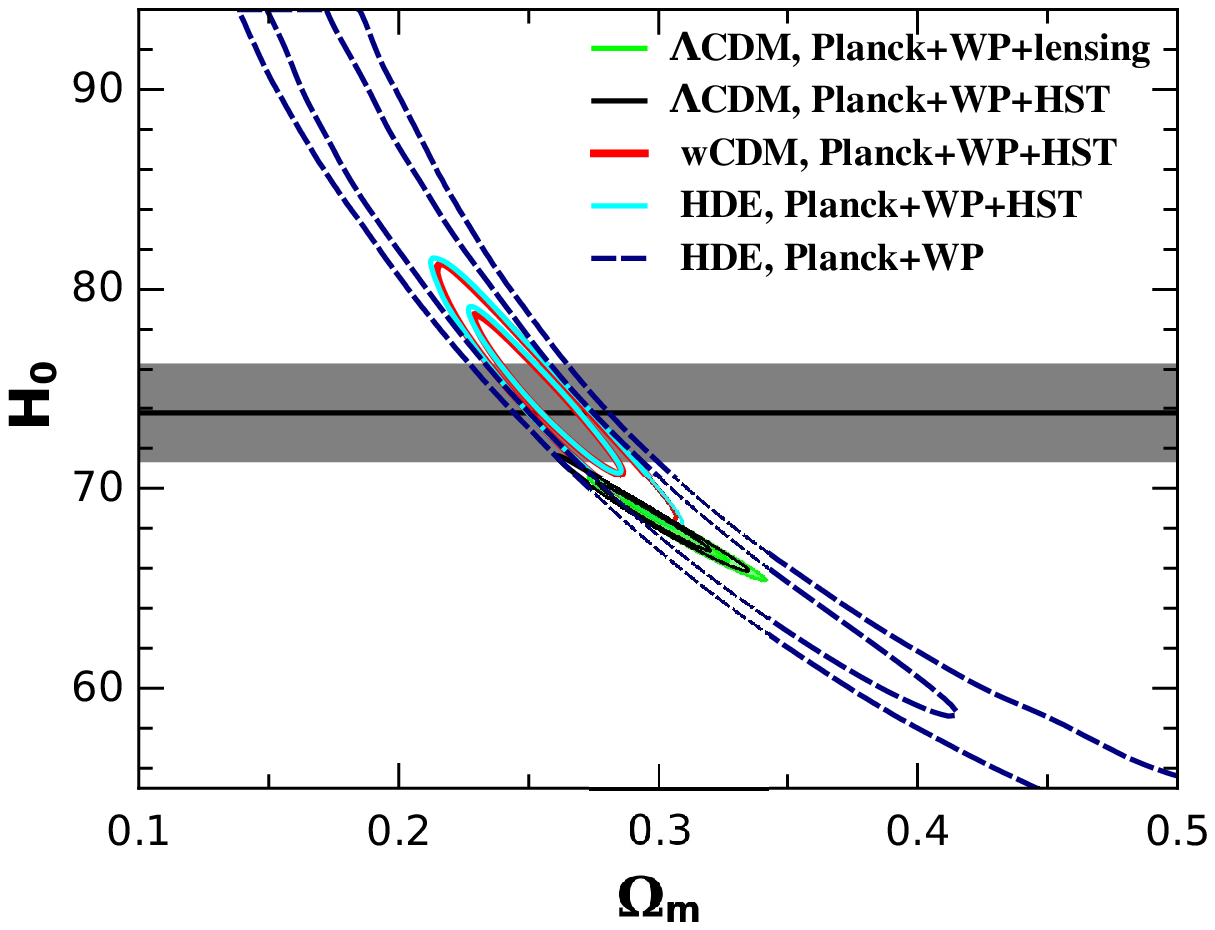}
\includegraphics[width=8cm,height=6cm]{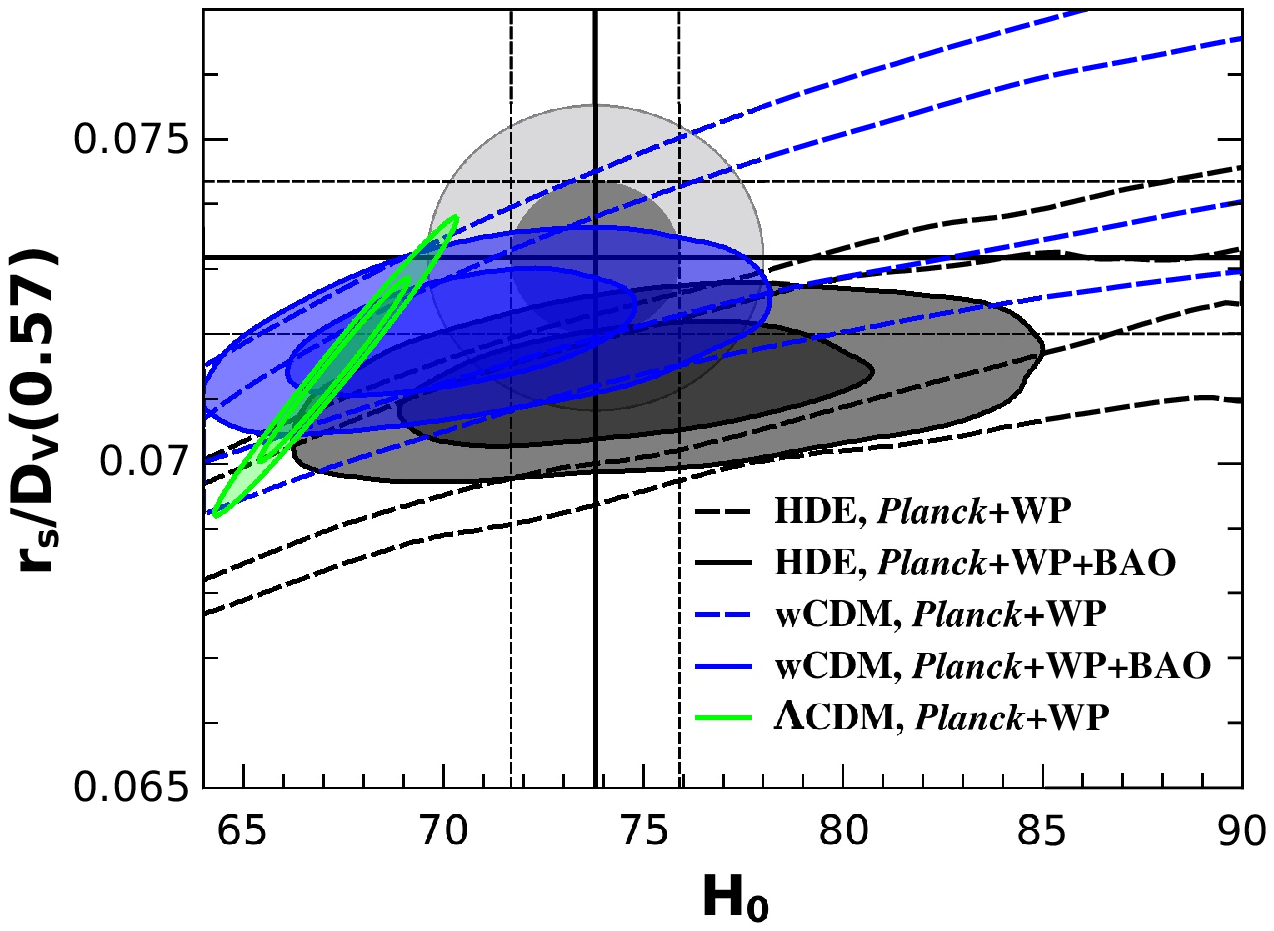}
\caption{\label{Fig:cmbbaohst_H0} Upper panels: Marginalized
likelihood distributions of $H_0$, for the $\Lambda$CDM (left
panel), $w$CDM (middle panel) and HDE (right panel) models. The
gray filled region represents the {\it HST} measurement result,
$H_0=73.8\pm 2.4$. Lower-left panel: Marginalized 68\% and 95\% CL
contours in the $\Omega_{\rm m}$--$H_0$ plane. The gray band shows
the result with $1\sigma$ range of the {\it HST} measurement.
Lower-right panel: Marginalized 68\% and 95\% CL contours in the
$H_0$--$r_{\rm s}/D_{\rm V}(0.57)$ plane. The horizontal solid and
dashed lines mark the central value and $1\sigma$ region of the
BOSS DR9 measurement, while the vertical lines mark the
observational result of the {\it HST}. The joint $1$ and $2\sigma$
likelihood region for BOSS DR9 + {\it HST} measurements is
represented by the dark and light gray shaded contours. }
\end{figure}

In the last section, we find that in the $w$CDM and HDE models the
CMB-only constraints allow a wide range of $H_0$ (see
Fig.~\ref{Fig:H0cmb}). Now, let us see if the tension between CMB
and {\it HST} can be relieved in these two models when the
external astrophysical data are added in the analysis. In the
upper panels of Fig.~\ref{Fig:cmbbaohst_H0} we plot the likelihood
distributions of $H_0$ in the $\Lambda$CDM (left), $w$CDM (middle)
and HDE (right) models, obtained by using {\it Planck}+WP (green
solid), {\it Planck}+WP+lensing (green dashed), {\it
Planck}+WP+BAO (red solid), {\it Planck}+WP+{\it HST} (blue
solid), and {\it Planck}+WP+BAO+{\it HST} (black solid),
respectively. In all plottings, the {\it HST} measurement result,
$H_0=73.8\pm 2.4$~\cite{HSTWFC3}, is shown in the gray filled
region. In the upper-left panel, we see that the constraints on
$H_0$ are fairly tight in the $\Lambda$CDM model, and the results
of various combinations involving {\it Planck}+WP are all in
tension with the {\it HST} measurement. However, for the $w$CDM
and HDE models, all the CMB combined constraints overlap well with
the gray region, showing that the tension between CMB and {\it
HST} is effectively relieved in these two models if the {\it
Planck} data are combined with BAO or/and {\it HST}. This
phenomenon can be seen more clearly in the $\Omega_{\rm m}$--$H_0$
plane for the three models (the lower-left panel). We see that the
allowed parameter space of the $\Lambda$CDM model is tightly
confined by the CMB data, and the positions of {\it
Planck}+WP+lensing (green solid) and {\it Planck}+WP+{\it HST}
(black solid) contours evidently deviate from the {\it HST}
measurement (the gray band). On the other hand, the CMB data alone
cannot effectively constrain the $\Omega_{\rm m}$--$H_0$ parameter
space for HDE (dark blue dashed). The positions of {\it
Planck}+WP+{\it HST} contours for the HDE (light blue solid) and
$w$CDM (red solid) models are all well consistent with the {\it
HST} measurement.

\begin{table}[!htp]
\caption{\label{Tab:dchi2} Residual $\chi^2$ values in the
$\Lambda$CDM, $w$CDM and HDE models}
\begin{center}
\label{table1}
\begin{tabular}{|c||c|c|c|}
  \hline
~~Model~~    &         ~  $\chi^2_{Planck{\rm+WP+BAO}} - \chi^2_{Planck{\rm+WP}}$  ~
&~ $\chi^2_{Planck{\rm+WP}+{\it HST}} - \chi^2_{Planck{\rm+WP}}$~
&  ~$\chi^2_{Planck{\rm +WP+BAO}+{\it HST}} - \chi^2_{Planck{\rm +WP}}$~ \\
 \hline
 $\Lambda$CDM & 2.5 & 7.8    &   9.1 \\
 \hline
  $w$CDM          & 2.6   & 1.0   & 3.7  \\
 \hline
  HDE  & 1.9   & 0.3  &  1.9 \\
 \hline
\end{tabular}
\end{center}
\end{table}

The tension between CMB and the external data sets (e.g., BAO and
{\it HST}) in the HDE model can be characterized by the $\Delta
\chi^2$ values, as listed in the last column of
Table~\ref{Tab:CMBExt}. The results are $\Delta \chi^2=1.7$, 0.3,
0.9, 0.9, and 0.2 for {\it Planck}+WP+BAO, {\it Planck}+WP+{\it
HST}, {\it Planck}+WP+BAO+{\it HST}, {\it WMAP}-9+BAO, and {\it
WMAP}-9+{\it HST}, respectively. These values are small, showing
that there is no severe tension between the data sets in the HDE
model. As a comparison with the $w$CDM and $\Lambda$CDM models, in
Table~\ref{Tab:dchi2} we show the residuals $\chi^2$ values of
{\it Planck}+WP+BAO, {\it Planck}+WP+{\it HST} and {\it
Planck}+WP+BAO+{\it HST} with respect to {\it Planck}+WP in the
three models. For the $\Lambda$CDM model, adding {\it HST} and
BAO+{\it HST} significantly increases the $\chi^2$ value by 7.8
and 9.1. The increments are 1.0 and 3.7 for the $w$CDM model, and
only 0.3 and 1.9 for the HDE model. Thus, the tension with {\it
HST} measurement is effectively relieved in the two dynamic dark
energy models.


Moreover, Fig.~\ref{Fig:cmbbaohst_H0} and Table~\ref{Tab:dchi2}
show that there is a better consistency among data sets in the HDE
model than in the $w$CDM model. The best-fit values of $H_0$ from
{\it Planck}+WP+BAO are 67.63, 69.68 and 72.63 for the
$\Lambda$CDM, $w$CDM and HDE models, among which the HDE result is
the most close to the {\it HST} measurement. To understand {\it
why {\it Planck}+WP+BAO gives a higher $H_0$ in the HDE model than
in the $w$CDM model}, in the lower-right panel of
Fig.~\ref{Fig:cmbbaohst_H0} we plot the $H_0$--$r_{\rm s}/D_{\rm
V}(0.57)$ contours for the three models. In this figure, we also
show the joint 1 and 2$\sigma$ likelihood region for BOSS DR9 +
{\it HST} measurements in the dark and light gray shaded contours.
We see that, for the $\Lambda$CDM model, the {\it Planck}+WP
contours (green solid) are consistent with the BOSS DR9
measurement, but are in tension with the {\it HST} measurement. In
the $w$CDM and HDE models, the allowed parameter spaces are
greatly broadened, and their {\it Planck}+WP contours (dashed
lines) overlap with the gray contours. Interestingly, the
positions of the $w$CDM and HDE contours are different: the HDE
contours lie in the smaller $r_{\rm s}/D_{\rm V}(0.57)$ region,
below the $w$CDM contours, so they overlap with the gray contours
at higher $H_0$ region. This helps us to understand why the {\it
Planck}+WP+BAO (black filled region) result of the HDE model has
higher values of $H_0$ than the {\it Planck}+WP+BAO (blue filled
region) result of the $w$CDM model.

Besides, it should be mentioned that, due to the anti-correlation
between $w$ (or $c$) and $H_0$, the {\it Planck}+WP+{\it HST} leads
to phantom results in the $w$CDM and HDE models. In \cite{Planck16},
the {\it Planck} Collaboration reported a result
$w=-1.24_{-0.19}^{+0.18}$ (95\% CL, {\it Planck}+WP+highL+BAO+{\it
HST}) for the $w$CDM model, which is in tension with $w=-1$ at the
more than 2$\sigma$ level. For the HDE model, the lower-left panel
of Fig.~\ref{Fig:cmbbaohst} shows that the 95\% CL contour from {\it
Planck}+WP+{\it HST} (red filled region) lies below the $z=0.5$ phantom
divide line (red dashed).

\subsection{Combined with SNIa}

In this subsection, we discuss the SNIa combined fitting results.

\begin{figure}[H]
\centering{
\includegraphics[width=8cm,height=6cm]{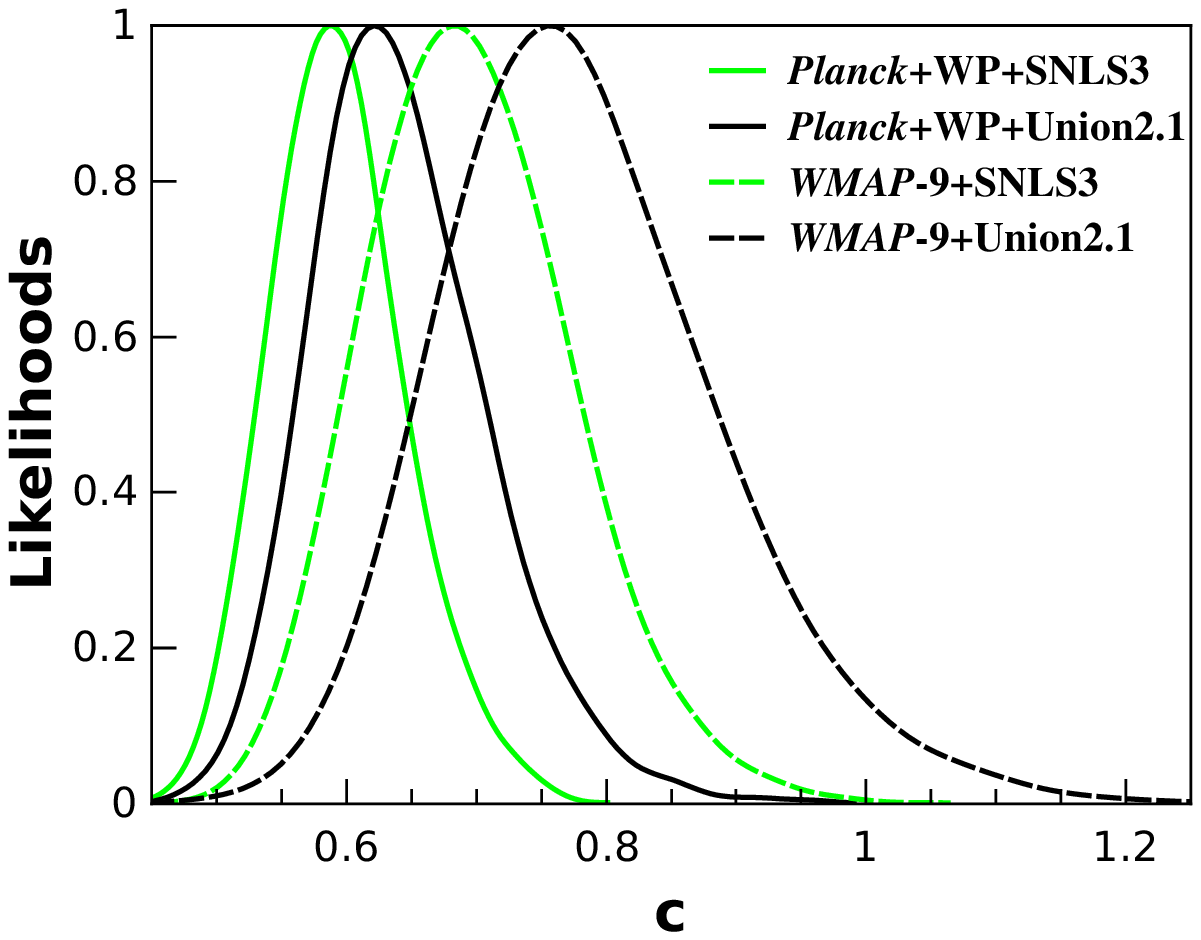}
\includegraphics[width=8cm,height=6cm]{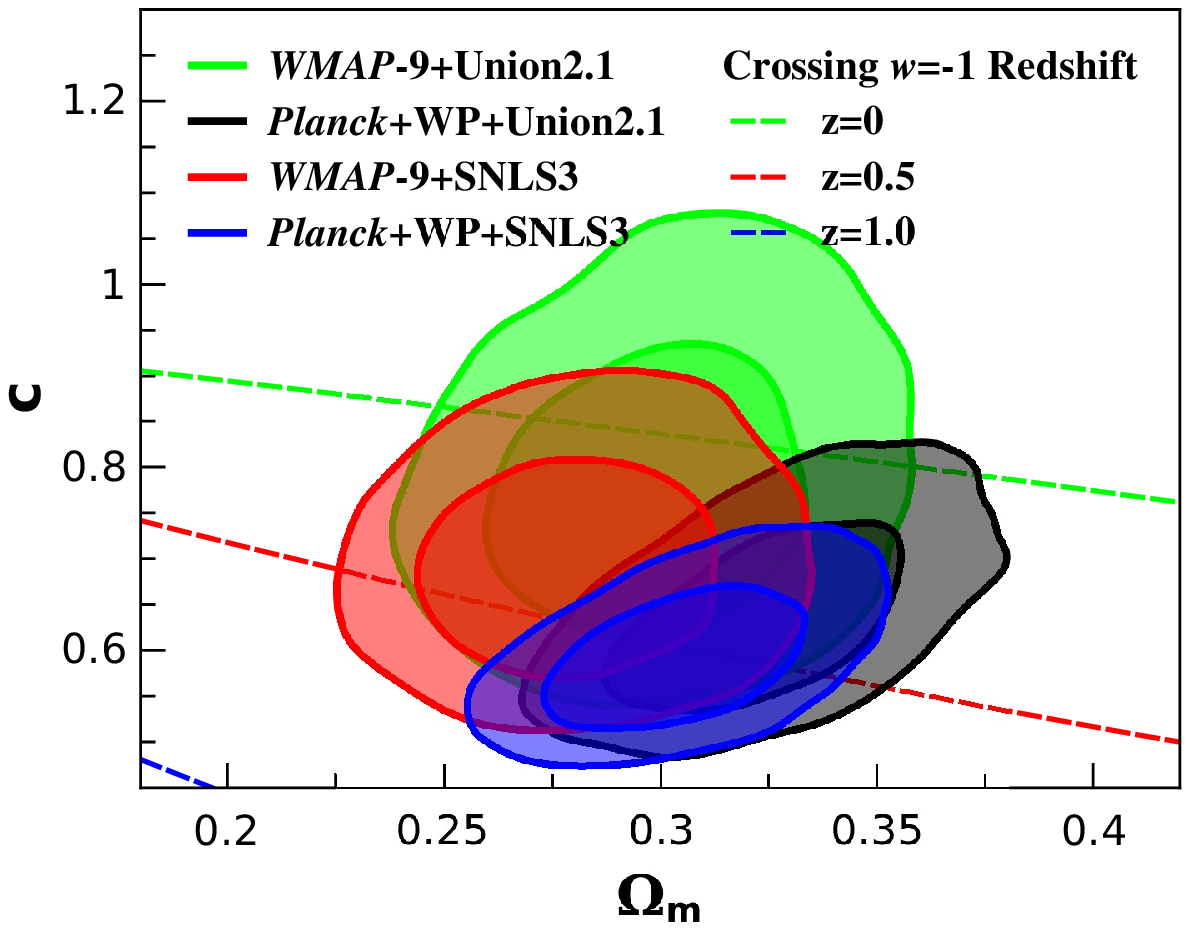}
\includegraphics[width=8cm,height=6cm]{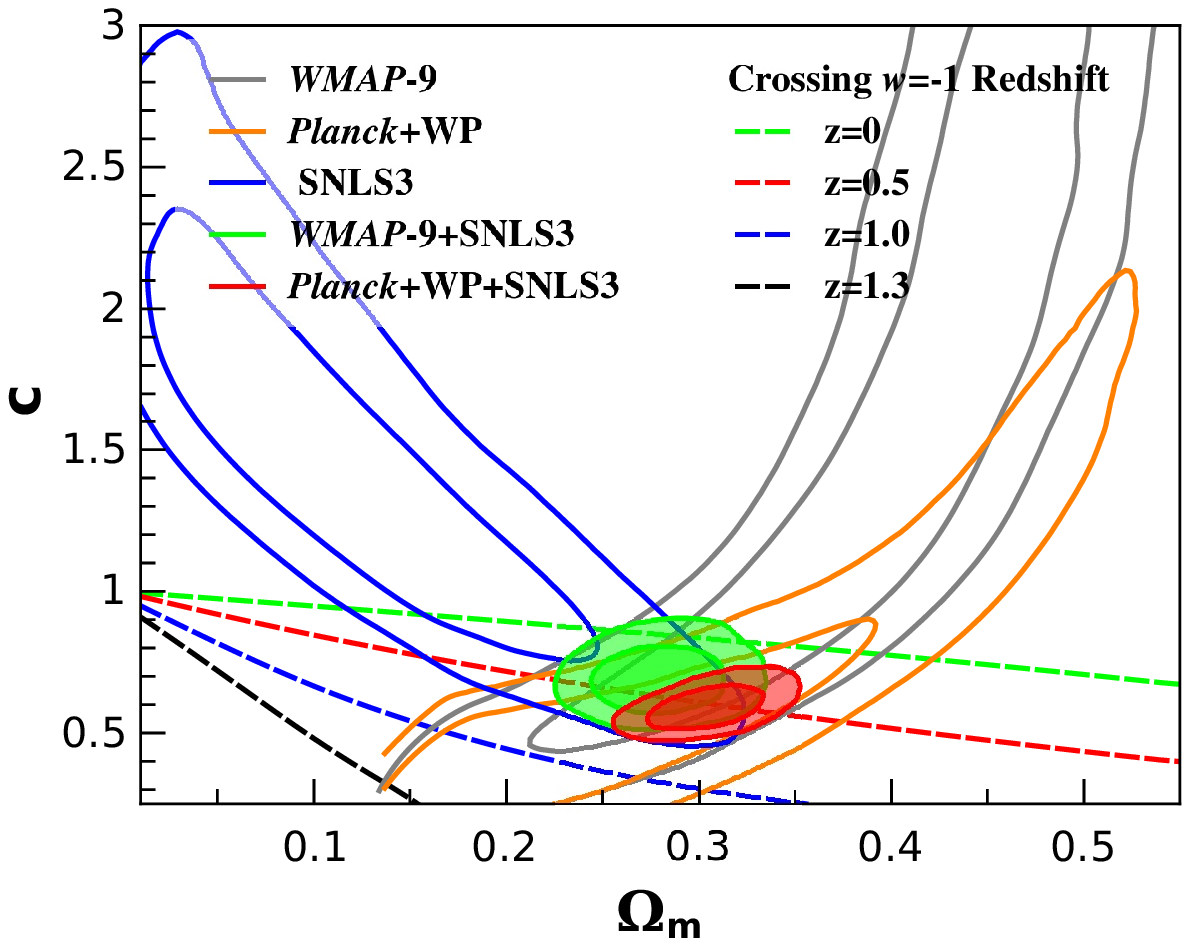}
\includegraphics[width=8cm,height=6cm]{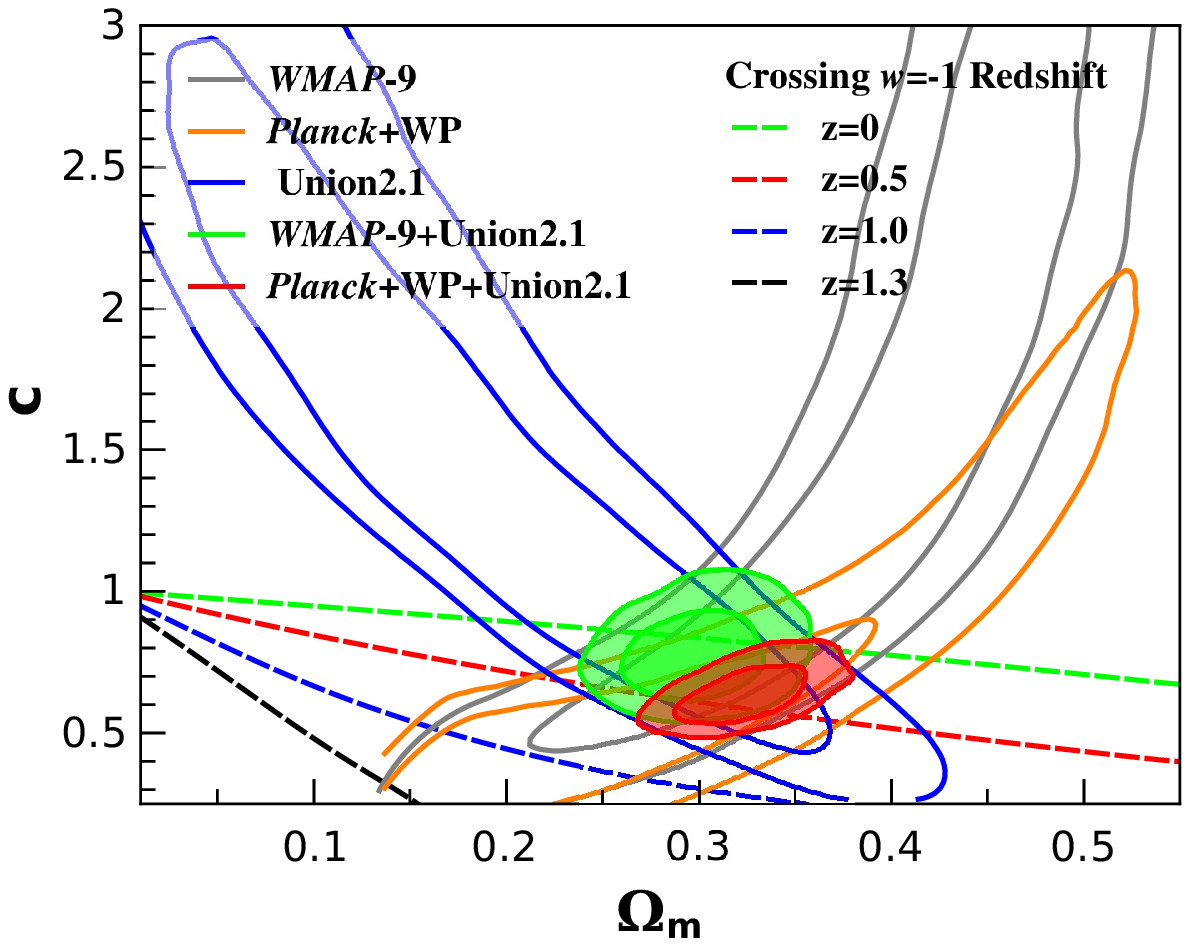}}
\caption{\label{Fig:cmbsn} Fitting results of the HDE model, from
CMB combined with BAO and {\it HST}. The upper-left panel shows
the marginalized distributions of $c$. The other three panels show
the marginalized 68\% and 95\% CL contours in the $\Omega_{\rm
m}$--$c$ plane, including the CMB+SNLS3 results (lower-left), the
CMB+Union2.1 results (lower-right), and a comparison of the
CMB+SNLS3 and CMB+Union2.1 results (upper-right).}
\end{figure}

The CMB+SNIa fitting results are plotted in Fig.~\ref{Fig:cmbsn}.
The likelihood distributions of $c$ are shown in the upper-left
panel. At the 68\% CL, we get $c=0.594\pm0.051$, $c=0.642\pm0.066$,
$c=0.696\pm0.078$ and $c=0.782\pm0.105$ for {\it Planck}+WP+SNLS3,
{\it Planck}+WP+Union2.1, {\it WMAP}-9+SNLS3 and {\it
WMAP}-9+Union2.1, respectively. Similar as the above results,
compared with the {\it WMAP}-9 results, the {\it Planck} results
have smaller best-fit values and error bars.
Adding lensing into the analysis effectively tightens the constraint,
yielding $c=0.583\pm0.042$ and $c=0.645\pm0.063$ for
{\it Planck}+WP+lensing combined with SNLS3 and Union2.1.
Compared with CMB+Union2.1, we find that CMB+SNLS3 yields more
phantom-like result.

In \cite{Planck16}, the {\it Planck} Collaboration reported that
there exists some tension between {\it Planck} and supernovae data
sets, and the tension between {\it Planck} and SNLS3 is more
severe than that between {\it Planck} and Union2.1. To investigate
the tension between CMB and SNIa data sets in the HDE model, in
the lower panels we plot the 68\% and 95\% CL contours in the
$\Omega_{\rm m}$--$c$ plane from {\it Planck}+WP (orange), {\it
WMAP}-9 (gray), SNIa (blue), {\it Planck}+WP+SNIa (red filled) and
{\it WMAP}-9+SNIa (green filled). The SNLS3 plottings are shown in
the lower-left panel, and the Union2.1 plottings are shown in the
lower-right panel. From the positions of the contours, we see that
the CMB data are consistent with Union2.1, but in tension with
SNLS3 (the 1$\sigma$ contours of CMB and SNIa do not overlap).
Table~\ref{Tab:CMBExt} shows that $\Delta \chi^2_{\rm {\it
Planck}+WP+SNIa}$, $\Delta \chi^2_{\rm {\it
Planck}+WP+lensing+SNIa}$ and $\Delta \chi^2_{\rm {\it
WMAP}-9+WP+SNIa}$ are 6.4, 7.3 and 3.5 for SNLS3, respectively,
while only 1.6, 3.4 and 0.1 for Union2.1, respectively. Besides,
as mentioned above, the results in Table~\ref{Tab:CMBExt} also
show some tension between SNLS3 and BAO+{\it HST}: for SNLS3 we
have $\chi^2_{\rm SNIa+BAO+{\it HST}}-\chi^2_{\rm
SNIa}-\chi^2_{\rm BAO+{\it HST}}=4.1$, while for Union2.1 the
value is only 1.0. So, it is fairly remarkable that {\it for the
HDE model the SNLS3 data set is in weak tension with all other
data sets}.

Another interesting phenomenon is that, although there is no
severe tension when we combine Union2.1 with BAO+{\it HST} or {\it
Planck}+WP, evident tension appears when we combine all these data
sets together. Table~\ref{Tab:CMBExt} shows that $\Delta
\chi^2_{Planck\rm +WP+lensing+Union2.1+BAO+{\it HST}}=9.6$, as
large as $\Delta \chi^2_{Planck\rm +WP+lensing+SNLS3+BAO+{\it
HST}}$ (that is equal to 10.9). This tension mainly comes from the
discrepancy between the results of {\it Planck}+WP+lensing and
Union2.1+BAO+{\it HST}: we find that $\chi^2_{Planck\rm
+WP+Union2.1+BAO+{\it HST}+lensing}-\chi^2_{Planck\rm
+WP+lensing}-\chi^2_{\rm Union2.1+BAO+{\it HST}+lensing}=8.6$. The
fitting results of Union2.1+BAO+{\it HST} are $\Omega_{\rm
m}=0.326\pm0.030$, $c=0.633\pm0.086$ and $H_0=73.09\pm2.36$, while
for {\it Planck}+WP+lensing the results are $\Omega_{\rm
m}=0.248\pm0.079$ and $c=0.508\pm0.207$. When we combine them, we
get $\Omega_{\rm m}=0.281\pm0.012$, $c=0.577\pm0.039$ and
$H_0=70.68\pm1.40$. These three sets of results do not match with
each other. Especially, the constraint result of $H_0$ in the
all-combined analysis is in tension with the {\it HST}
measurement.

For {\it WMAP}-9 we find that $\Delta \chi^2_{\rm {\it
WMAP}-9+SNIa+BAO+{\it HST}}=5.6$ and 4.3 for SNLS3 and Union2.1,
respectively, which also implies some tension, but not so severe
as the {\it Planck} case. Thus, it is no longer viable to do a
all-combined analysis by combining {\it Planck} data with all the
external data sets of SNIa, BAO and {\it HST}. Our tightest {\it
self-consistent} constraint is $c=0.495\pm0.039$ obtained from
{\it Planck}+WP+BAO+{\it HST}+lensing.

\section{Concluding remarks}\label{sec:concl}

In this paper we perform detailed investigation on the constraints
on the HDE model by using the {\it Planck} data. We find the
following results:

\begin{itemize}
\item HDE provides a good fit to the {\it Planck} high-$\ell$
temperature power spectrum. The discrepancy at $\ell\lesssim
20-40$ found in the $\Lambda$CDM model remains unsolved in the HDE
model. The best-fit power spectra of the $\Lambda$CDM, $w$CDM and
HDE models are similar to each other at $\ell\gtrsim25$. In the
$\ell\lesssim25$ region, the $w$CDM and HDE spectra have slightly
lower amplitudes than the $\Lambda$CDM spectrum.

\item {\it Planck} data alone can lead to interesting constraint on $c$.
By using {\it Planck}+WP+lensing, we get $c=0.508\pm0.207$ (68\%
CL), favoring the present phantom behavior of HDE at the more than
2$\sigma$ CL. Comparably, by using {\it WMAP}-9 data alone we cannot
get valuable constraint on $c$.

\item In the HDE model, we find $A_{\rm L} > 1$ at the 2.2$\sigma$ and 1.7$\sigma$ levels 
by using the {\it Planck}+WP and {\it Planck}+WP+lensing data.
So, HDE cannot help remove or relieve the anomaly of $A_{\rm L}$ 
(i.e., the preference for high $A_{\rm L}$ in the temperature power spectrum).

\item At the 68\% CL, the results are $c=0.484\pm0.070$,
$c=0.474\pm0.049$, $c=0.594\pm0.051$, and $c=0.642\pm0.066$ from
{\it Planck}+WP combined with BAO, {\it HST}, SNLS3 and Union2.1,
respectively. The constraints can be improved by 2\%--15\% if we
further add the {\it Planck} lensing data into the analysis. The
results from {\it WMAP}-9 combined with each Ext are
$c=0.746\pm0.165$, $c=0.569\pm0.086$, $c=0.696\pm0.078$ and
$c=0.782\pm0.105$. Compared with {\it WMAP}-9+Ext results, we find
that {\it Planck}+WP+Ext results reduce the error by 30\%--60\%,
and prefer a more phantom-like HDE.

\item Non-standard dark energy models are helpful in relieving the
tension between CMB and {\it HST} measurements. In the CMB-only
analysis, the strong correlation between $c$ ($w$) and
$\Omega_{\rm m} h^3$ in the HDE ($w$CDM) model makes $H_0$
unconstrained. We find that $\chi^2_{Planck\rm +WP+{\it
HST}}-\chi^2_{Planck\rm +WP}=7.8$, 1.0 and 0.3 for the
$\Lambda$CDM, $w$CDM and HDE models, respectively.

\item There is no evident tension when we combine {\it Planck}+WP
with BAO, {\it HST} or Union2.1: values of $\Delta
\chi^2\equiv\chi^2_{Planck\rm +WP+Ext}-\chi^2_{Planck\rm
+WP}-\chi^2_{\rm Ext}$ for them are 1.7, 0.3 and 1.6,
respectively. The SNLS3 data set is in weak tension with the other
data sets. When SNLS3 is combined with {\it Planck}+WP, {\it
Planck}+WP+lensing, {\it WMAP}-9 and BAO+{\it HST}, we obtain
large values of $\Delta \chi^2$, equal to 6.4, 7.3, 3.5 and 4.1,
respectively.

\item The {\it Planck}+WP+BAO and {\it Planck}+WP+{\it HST} results are in good agreement with each other.
The best-fit and 68\% CL constraints on $H_0$ in the {\it
Planck}+WP+BAO analysis are $H_0=72.63$ and $H_0=75.06\pm3.82$,
close to the {\it HST} measurement result, $H_0=73.8\pm 2.4$.

\item Although Union2.1 is not in tension with CMB or BAO+{\it HST},
the combination Union2.1+BAO+{\it HST} is in tension with the
combination {\it Planck}+WP+lensing. When we combine the two
together, we find $\Delta \chi^2=8.6$. So it is not viable to do an
all-combined analysis for HDE by using the {\it Planck} data
combined with all the Exts. Our tightest self-consistent constraint
is $c=0.495\pm0.039$ obtained from {\it Planck}+WP+BAO+{\it
HST}+lensing.
\end{itemize}

\begin{acknowledgments}
We acknowledge the use of {\it Planck} Legacy Archive and the
discussion with Gary Hinshaw. We thank KIAS Center for Advanced
Computation and Institute for Theoretical Physics for providing
computing resources.
XDL thanks Juhan Kim for valuable discussions and kind help.
ZHZ thanks Cheng Cheng for kind help.
ML and ZZ are supported by the National
Natural Science Foundation of China (Grant Nos.~11275247 and
10821504). XDL is supported by the Korea Dark Energy Search (KDES)
grant. YZM is supported by the Natural Sciences and Engineering
Research Council of Canada and Canadian Institute for Theoretical
Astrophysics (CITA). XZ is supported by the National Natural
Science Foundation of China (Grant Nos.~10705041, 10975032 and
11175042), and by the National Ministry of Education of China
(Grant Nos.~NCET-09-0276, N100505001 and N120505003).
\end{acknowledgments}


\end{document}